\documentclass[a4paper,groupedaddress,superscriptaddress,twocolumn]{revtex4-2} 

\usepackage{graphicx}
\usepackage{physics}
\usepackage[usenames,dvipsnames]{xcolor}
\usepackage[normalem]{ulem}
\usepackage{rotating}
\usepackage{subcaption}

\usepackage{amsmath}
\usepackage{bm}

\newcommand{\vect}[1]{\bm{#1}}  

\newcommand{\hatF}[1]{\tilde{#1}}  

\usepackage{url}
\usepackage{hyperref}

\usepackage{multirow}

\hypersetup{colorlinks=true,breaklinks,linkcolor=blue,urlcolor=blue,citecolor=green}

\begin{document}
\title{Theory of x-ray absorption spectroscopy for ferrites
}

\date{\today}

\author{Felix Sorgenfrei}
\affiliation{Department of Physics and Astronomy, Uppsala University, Box-516,Uppsala  SE-751 20 Sweden}
\author{M\'ebarek Alouani}
\affiliation{Université de Strasbourg, IPCMS, CNRS-UNISTRA UMR 7504, 67034 Strasbourg, France} 
\author{Johan Sch\"ott}
\affiliation{Department of Physics and Astronomy, Uppsala University, Box-516,Uppsala  SE-751 20 Sweden}
\affiliation{Orexplore AB, Torshamngatan 30B, 164 40 Kista, Stockholm, Sweden}
\author{H. Johan M. J\"onsson}
\affiliation{Department of Physics and Astronomy, Uppsala University, Box-516,Uppsala  SE-751 20 Sweden}
\affiliation{School of Computational Sciences, Korea Institute for Advanced Study, Seoul 02455, South Korea}
\author{Olle Eriksson}
\affiliation{Department of Physics and Astronomy, Uppsala University, Box-516,Uppsala  SE-751 20 Sweden}
\affiliation{Wallenberg Initiative Materials Science, Uppsala University, Box-516,Uppsala  SE-751 20 Sweden}
\author{Patrik Thunstr\"om}
\email{patrik.thunstrom@physics.uu.se}
\affiliation{Department of Physics and Astronomy, Uppsala University, Box-516,Uppsala  SE-751 20 Sweden}

\begin{abstract}
The theoretical calculation of the interaction of electromagnetic radiation with matter remains a challenging problem for contemporary \textit{ab initio} electronic structure methods, in particular for x-ray spectroscopies. This is not only due to the strong interaction between the core-hole and the photo-excited electron, but also due to the elusive multiplet effects that arise from the Coulomb interaction among the valence electrons.
In this work we report a method based on density-functional theory in conjunction  with multiplet ligand-field theory to investigate various core-level spectroscopies, in particular x-ray absorption spectroscopy (XAS) and x-ray magnetic circular dichroism (XMCD). 
The developed computational scheme is applied to the $L_{2,3}$ XAS edges  of magnetite (Fe$_3$O$_4$) as well as
cobalt ferrite (CoFe$_2$O$_4$) and nickel ferrite (NiFe$_2$O$_4$)
and the corresponding XMCD spectra. The results are in overall good agreement with experimental observations, both  regarding the XAS $L_2$/$L_3$ branching ratio, the peak positions as well as the relative intensities. The agreement between theory and experiment is equally good for XAS and the XMCD spectra, for all studied systems. The results are analyzed in terms of $e_g$ and $t_{2g}$ orbitals contributions and the importance of optimizing the Slater parameters. The analysis also highlights the strong effect  of the $2p$-$3d$ interaction in x-ray spectroscopy.
\end{abstract}


\maketitle

\section{Introduction}
Core-level spectroscopy is one of the most powerful techniques for probing the electronic structure of strongly correlated electron systems, as 
the interaction between x-rays and matter gives element-specific insights into the magnetic and electronic properties of materials, as well as their structural and chemical properties. In particular, the development of synchrotron radiation facilities and recently, x-ray free electron lasers, have made the analysis efficient and the resolution very high in energy and in specific cases, in time~\cite{Altarelli}.
This increase in the experimental resolution has made it possible to detect fine features and structures in the x-ray spectra~\cite{Glatzel200565,PhysRevLett.111.253002}.
This development of experimental tools and techniques calls for a novel theoretical approach to compute 
x-rays absorption spectra, most notably, for systems with complex or correlated electronic structures, like magnetites as well as 
cobalt- and nickel ferrites. In this work, we focus specifically on these materials and on new insights for the theoretical description of x-ray absorption spectroscopy (XAS) and x-ray magnetic circular dichroism (XMCD).

For the theoretical modelling of core-level and x-ray absorption spectroscopy, there have been two main distinct methodologies that are primarily  used today. 
The first one is based on density functional theory (DFT) and the Kubo formula~\cite{RevModPhys.71.1253} while the second focuses on multi configuration effects~\cite{cowan-multiplets}. 
Both methodologies are well reviewed in the literature (see for example Ref.~\cite{deGroot-book,yaresko-book}).
The DFT-based methods are essentially parameter-free and have, as a main advantage, the description of magnetism, structural properties and chemical bonding in general. These methods are designed for the ground-state properties of materials, and hence their description of the excited states probed with x-ray spectroscopy is very limited. This is due to an incomplete treatment of electronic correlations, in particular the multiconfigurational effects. Overall, these methods work best for widely dispersive electron systems. In contrast,
atomic multiplet theory is essentially intended for a better description of the multiconfigurational effects originating from on-site electron-electron interaction~\cite{MLFT}. For practical computational efficiency, the multiplet theory is often implemented for an isolated ion and thus neglects hybridization effects. It is therefore not surprising that this implementation of the theory has been found to work best for localized electronic states, e.g., the $4f$ orbitals of rare-earth elements. However, effects like crystal-field splitting and orbital hybridization can be treated by adding suitable model terms to the Hamiltonian of the system. The resulting theory is often referred to as the multiplet ligand-field theory (MLFT)~\cite{MLFT}. This theory is very powerful in reproducing experimental spectra when the electrons have a tendency to be localized, but has the obvious drawback in the ambiguity of choosing and justifying the various parameters of the model Hamiltonian (see e.g. Ref.~\cite{1742-6596-190-1-012143}). One way to reduce the ambiguity is to obtain the hybridization strength parameters and the crystal-field splitting from a tight-binding parameterization of a DFT calculation~\cite{PhysRevB.85.165113}. For the $4f$-orbitals of rare-earth systems it is also possible to neglect the hybridization and extract the remaining MLFT parameters completely from DFT~\cite{Ramanantoanina2019}. 

To describe both on-site electron-electron interaction and hybridization effects, many methods have been proposed, for example DFT+$U$~\cite{PhysRevB.44.943} and DFT in conjunction with dynamical mean-field theory (DFT+DMFT)~\cite{RevModPhys.78.865}.
Since DFT+DMFT takes into account the multiconfigurational effects, it was shown to provide a better description of valence-band spectral function of, e.g., the 3$d$ transition-metal oxides (TMOs)~\cite{PhysRevB.74.195114,PhysRevLett.109.186401}. 

In the x-ray absorption process, the excitation involves the promotion of an electron in a core orbital to an empty orbital, according to the selection rules of the electric dipole operator~\cite{DEGROOT1993111}. The resulting core-hole interacts with the valence electrons and the photo-electron. For example, for the $L_{2,3}$-edges of transition metal complexes the presence of a $2p$ core-hole in the final state tends to further localize the valence states. In the final-state approximation, the core hole is treated as an attractive static potential, and the valence electrons and the photo-electron are allowed to relax in the presence of this potential. However, it was shown that applying the final state approximation in DFT+DMFT does not improve the calculated XAS spectra of Fe and Co, and in the case of Ni dramatically worsens the comparison with the experiment~\cite{PhysRevB.84.115102}. Several propositions to include the dynamics of the core-hole  within a single computational framework for calculating $L$-edges of transition metal (TM) systems are available in the  literature, such as the multiple scattering approach~\cite{Nesvizhskii:nl4402} with dynamical screening from time-dependent DFT~\cite{PhysRevLett.80.4586,PhysRevB.67.115120},  
the application of configuration interaction to a set of molecular orbitals~\cite{ikeno-ci} and the Bethe-Salpeter equation~\cite{PhysRevB.82.205104,PhysRevB.83.115106}.
In spite of a high level of theoretical complexity, most of the methods mentioned above do not deliver a sufficiently accurate description of XAS of a wide range of materials, exemplified by the transition metal oxides. In recent years, approaches based on extending the impurity model in DFT+DMFT with core states have also been used to simulate core level x-ray photoemission spectroscopy~\cite{Hariki2017} and resonant x-ray emission spectroscopy~\cite{ Kolorenc2018}. A good overview of most methods for calculating x-ray absorption spectra is presented in~\cite{DeGroot2021}.

In the present study we use MLFT in conjunction with DFT+$U$ and DFT+DMFT to describe on a equal footing the multiconfigurational effects as well as the dynamics of the core-hole photo-electron interaction and its effect on XAS and XMCD. The method described here has key components that are identical for MLFT combined with DFT+$U$ and DFT+DMFT, and so for simplicity we have chosen to give most numerical examples from the MLFT+DFT+$U$ approach. The aim of the present theory is to rely on as few free parameters as possible. For example, the hybridization of the orbitals is directly extracted from the electronic structure calculation and the screening of the local electron-electron interaction (Hubbard $U$) is estimated using constrained DFT (cDFT). However, there currently is no reliable way to calculate the screening of the zeroth order valence-core interaction ($F_{pd}^0$), so it is treated as a free parameter. We also investigate how sensitive the spectra are to the description of the screening of the higher-order Slater parameters.
In this work transition selection rules have been included within the electric-dipole approximation. We present the details of the implementation and the results from an application of the developed method to the description of the $L_{2,3}$-edges and XMCD spectra of magnetite, Fe$_3$O$_4$, as well as cobalt- (CoFe$_2$O$_4$) and nickel ferrite (NiFe$_2$O$_4$) at room temperature in the inverse spinel structure. 
We show that our theoretical results compare favorably well with experimental observations. In particular, we show that both the $L_3$/$L_2$ experimental XAS branching ratio, as well as fine structures in the XAS and XMCD spectra, are well reproduced by our calculations, contrary to previous DFT calculations. 

\section{Theory}
\label{sec:theory}

\subsection{DFT+MLFT} 
\label{DFT+MLFT}
In the following, we describe the DFT+MLFT approach, and show how to use this theory to compute x-ray absorption spectra and x-ray magnetic circular dichroism of the magnetite Fe$_3$O$_4$  as well as the cobalt and nickel ferrites. The theoretical model can in a simplified way be described as a transfer of parameters obtained from DFT (or DFT+$U$) based methods, to a MLFT level of theory. In this way, hybridization effects and crystal-field splittings do not enter the theory as fitting parameters but rather evaluated from \textit{ab initio} theory. We describe the essential aspects of this mapping below.  

\subsubsection{Projection}
\label{projection}
The starting point of the theory involves the one-particle DFT+$U$ Green's function of the lattice, which encodes the contributions from hybridization and crystal-field splittings of the transition metal $3d$ orbitals. It is defined as the resolvent of the lattice-momentum dependent Hamiltonian $\tilde{H}_k^\text{DFT}$ and the static DFT+$U$ self-energy $\tilde{\Sigma}_{k}$;
\begin{eqnarray}
\tilde G_{k}(\omega)  = ((\omega + \mu)\tilde{1} - \tilde{H}_k^\text{DFT} -\tilde{\Sigma}_k)^{-1},
\label{eqn1}
\end{eqnarray}
where $\mu$ is the chemical potential and $\tilde{1}$ is the identity operator. The self-energy $\tilde{\Sigma}_{k}$ is obtained from the local self-energy $\tilde{\Sigma}^{\mathrm{loc}}$ and a double-counting correction $\tilde{\Sigma}^{\mathrm{loc}}_{\mathrm{DC}}$ through a projection from a set of impurity orbitals to the lattice orbitals:
\begin{eqnarray}
\tilde{\Sigma}_k = \tilde{P}_{k} (\tilde{\Sigma}^{\mathrm{loc}} - \tilde{\Sigma}^{\mathrm{loc}}_{\mathrm{DC}}) \tilde{P}_{k}^{\dagger},
\label{eq:sigma}
\end{eqnarray}
where $\tilde{P}_{k}$ is the projection operator from the orthonormal orbitals on the impurity site to the lattice orbitals at lattice momentum $k$. The impurity orbitals are defined through the construction of the projection as detailed in Refs.~
\cite{PhysRevB.76.035107,PhysRevLett.109.186401,oscargranas}. In the construction of the projection operator we only use the DFT bands within an energy window $[-16,10]$ eV around the Fermi level, in order to preserve the TM 3$d$ character of the impurity orbitals. The lower bound of the energy window prevents the impurity orbitals to obtain any oxygen 2$s$ character, which adversely affects the crystal-field splitting in the impurity orbitals. The upper bound prevents the impurity orbitals to include high-energy TM 4$d$ and TM 5$d$ characters.

The local Green's function is obtained from $\tilde{G}_{k}(\omega)$ through the reverse projection:
\begin{eqnarray}
\tilde{G}^{\mathrm{loc}}(\omega) = \frac{1}{N_k}\sum_k^{N_k} \tilde{P}_{k}^{\dagger} \tilde{G}_{k}(\omega) \tilde{P}_{k},
\label{eq:Glocal}
\end{eqnarray}
where $N_k$ is the number of $k$-points in the first Brillouin zone.
The projection of the lattice Green's function to the local Green's function is identical to that in dynamical mean-field theory (DMFT).
The hybridization function 
\begin{eqnarray}
\tilde{\Delta}(\omega) = (\omega +\mu)\tilde{1}- \tilde{H}_0^\mathrm{loc}  -\tilde{\Sigma}^\mathrm{loc} + \tilde{\Sigma}^{\mathrm{loc}}_{\mathrm{DC}} -\tilde{G}^{\mathrm{loc}}(\omega)^{-1} 
\label{eq:hyb_operator}
\end{eqnarray}
describes the hybridization of the $3d$-TM orbitals with the orbitals of the rest of the material. The imaginary part of the hybridization function is similar to the density of states but has peaks at the energies of the hybridising orbitals. The intensity of these peaks corresponds to the strength of the hybridization. The role of the ligand orbitals is to mimic this $ab-initio$ hybridization i.e. to allow ligand electrons to enter (and leave) the $3d$-TM orbitals with the appropriate transition energy.
In Eq.~\eqref{eq:hyb_operator} the local Hamiltonian $\tilde{H}^\mathrm{loc}$ is calculated from $\tilde{H}_k^\text{DFT}$ by using the same projection procedure as in Eq.~\eqref{eq:Glocal}.

\subsubsection{Discretization of the hybridization function $\Delta(\omega)$ \label{appendix:hyb}}

The hybridization function operator in Eq.~\eqref{eq:hyb_operator} can be approximated as
\begin{eqnarray}
\tilde{\Delta}^{\mathrm{ED}}(\omega) = \tilde{V} \Big[ \omega\tilde{1} - \tilde{H}^{\mathrm{bath}}\Big]^{-1} \tilde{V}^{\dagger},
\label{eq:hyb_sum}
\end{eqnarray}
where $\tilde{H}^{\mathrm{bath}}$ is an effective Hamiltonian for a finite (usually small) set of auxiliary bath orbitals which hybridize with the impurity orbitals according to the impurity-bath hopping operator $\tilde{V}$. This approximation is routinely done in the exact diagonalization (ED) solver in DMFT~\cite{PhysRevLett.72.1545}. The matrix elements of the operators $\tilde{V}$ and $\tilde{H}^{\mathrm{bath}}$ are parameterized and the parameters are set to reproduce the main peaks of the imaginary part of the exact hybridization function $\tilde{\Delta}(\omega)$. In our implementation several energy windows are selected after an inspection of $\Delta(\omega)$, and within each window the parameters for one bath orbital is fitted to minimize the difference between $\Delta^{\mathrm{ED}}$ and $\Delta(\omega)$. The matrix representation $\Delta(\omega)$ of $\tilde{\Delta}(\omega)$ depends on the choice of impurity orbitals. $\Delta(\omega)$ becomes block diagonal if the impurity orbitals are symmetrized such that they transform according to the irreducible representations of the system. Even in the absence of global symmetries it is often advantageous to symmetrize the impurity orbitals according to their local environment to minimize the off-diagonal elements in the fitting of $\Delta(\omega)$.

\subsubsection{Impurity Hamiltonian}
The total impurity Hamiltonian for the TM $3d$ and $2p$ shells is summarized as
\begin{align}
\label{eq:totalXASHam}
\hat{H} = & \sum_{ij} H^\textrm{3d}_{ij}\hat{c}_i^\dagger \hat{c}_j + \sum_{bb'} H_{bb'}^\mathrm{bath}  \hat{c}_b^\dagger \hat{c}_{b'} + \sum_{i,b} (V_{i,b}\hat{c}^\dagger_i \hat{c}_b + \textrm{h.c.}) \nonumber \\
&+ \sum_{ij} H^\textrm{2p}_{ij}\hat{c}_i^\dagger \hat{c}_j   + \hat{H}_U,
\end{align}
with
\begin{align}
H^\textrm{3d} = & H^\mathrm{loc} - \Sigma^{\mathrm{loc}}_{\mathrm{DC}} + H^\textrm{SOC}_{3d},\\
H^\textrm{2p} = & \epsilon_p\tilde{1} + H^\textrm{SOC}_{2p},
\end{align}
where $H^\textrm{3d}$ and $H^\textrm{2p}$ contains the $3d$ and $2p$ on-site energies including the spin-orbit coupling (SOC) (see Eq.~\eqref{eq:XAS:SOC}), respectively. $H^\mathrm{bath}$ and $V$ describe the bath and its coupling to the impurity, and $\hat{H}_U$ is the Coulomb interaction (see Eq.~\eqref{eq:DMFT:Coulomb} of the appendix). 

The diagonalization of Eq.~\eqref{eq:totalXASHam} results in the set of many-body eigenstates $\ket{n}$, each expressed by a sum of product states, and the corresponding eigenenergies $E_n$:
\begin{eqnarray}
\hat H  \ket{n} = E_n \ket{n}.
\label{eq-Ei}
\end{eqnarray}
Once the eigenstates with eigenenergies at most a few $k_B T$ above the lowest eigenenergy are found using a Lanczos algorithm, we use these states to calculate spectra and other observables such as occupation numbers and spin moments.

\subsection{Spectra}  
Using the eigenstates $\ket{n}$ and eigenenergies $E_n$ from the Hamiltonian $\hat{H}$ in Eq.~(\ref{eq:totalXASHam}), we calculate the XAS and XMCD spectra at finite temperature (300 K) according to
\begin{eqnarray}
I (\omega) =   \frac{1}{Z} \sum_n  -\Im (G_n(\omega))  \exp(-\beta E_n),\label{eqn:spectra1}
\end{eqnarray}
with 
\begin{eqnarray}
G_n(\omega) =    \bra{n}\hat T^\dagger \frac{1}{(\omega + E_n +i\Gamma_c) \hat{1}- \hat{H} } \hat T \ket{n},\label{eqn:spectra2}
\end{eqnarray}
and where $\omega$ is the incoming photon energy, $\hat{1}$ is the identity operator, $Z$ the partition function and $\beta$ the inverse temperature. Here $\Gamma_c$ models the core-hole decay rate and defines the energy resolution in terms of the half-width at half-maximum (HWHM). The transition operator $\hat{T}$ is, within the dipole approximation, equal to $\hat{D} = \bm{\epsilon} \cdot \hat{\bm{p}}$, where $ \bm{\epsilon}$ is the light polarization vector and $\hat{\bm{p}}$ is the momentum operator.  
We have implemented Eqns.~(\ref{eqn:spectra1}) and (\ref{eqn:spectra2}) using a Lanczos algorithm in an open-source software~\cite{JohanSchott,GitHubImpurityModel}.

\section{Details of the calculations}
\label{details}
The Hamiltonian in Eq.~(\ref{eq:totalXASHam}) contains several parameters that either can be calculated or have been estimated. 
From the electronic structure calculation, we determine the local impurity Hamiltonian from a direct projection and also construct the hybridization function. The hybridization strength ($V$) and the bath state energies ($e_b$) are set to reproduce the hybridization function.
The average Coulomb repulsion between the electrons within the 3$d$ and 2$p$ orbitals of the TM atom can be expressed in terms of the Slater-Condon parameters $F^{k}_{l_il_i}$, $G^{k}_{l_il_j}$ and $F^{k}_{l_il_j}$ and the Wigner 3$j$ symbols as~\cite{cowan-multiplets}
\begin{align}
\label{eq:averageUii}
    U_{l_il_i} & =F^0_{l_il_i}-\frac{2 l_i+1}{4l_i+1}\sum_k \begin{pmatrix}l_i & k & l_i\\ 0& 0& 0 \end{pmatrix}^2 F^{k}_{l_il_i},\\
    U_{l_il_j} & =F^{0}_{l_il_j}-\frac{1}{2} \sum_k\begin{pmatrix}l_i & k & l_j\\ 0& 0& 0 \end{pmatrix}^2 G_{l_il_j}^{k},
\end{align}
where $l$ denotes the angular momentum of the orbitals. Notice that in these expressions $k$ is not a crystal momentum (as it is in Section \ref{projection}) but an angular momentum. The bare higher-order Slater parameters ($k \geq 1$) are calculated by solving the Slater-Condon integrals. To account for screening effects the parameters are then individually reduced to between 70\% and 90\% of their bare values. The values within this interval are chosen to best fit the experiment. For comparison, we also performed calculations where the higher-order Slater parameters are kept at a fixed ratio of 80\% of their bare values, to show the impact of the screening. The screening of the zeroth-order parameters, $F_{dd}^0$ and $F_{pd}^0$, are typically so high that they cannot be estimated by this approach. Therefore, we have calculated $F_{dd}^0$ using cDFT for NiFe$_2$O$_4$ using the electronic structure code Wien2K~\cite{Blaha2020}. There is currently no accurate method for calculating the screening of $F_{pd}^0$, so it was varied in the empirically motivated interval $F_{dd}^0+1\,$eV$\leq F_{pd}^0< 1.4 F_{dd}^{0}$.

The MLFT double counting of the $d$-orbitals is given by,
\begin{align}
    \Sigma_{DC}= n_{d} U_{dd}-n_{p}U_{dp}-\delta_{DC},
\end{align}
where $n_{d}$ is the occupation of the $d$ orbitals, $n_p$ is the occupation of the $p$ core states and $\delta_{DC}$ is the charge transfer potential. The value of $\delta_{DC}$ was taken from Ref.~\cite{PhysRevB.96.245131}.

Compared to the previous work the $p$-orbital potential $\Delta \epsilon_p$ was introduced to reproduce the relative difference in the core-level binding energies of the different Fe sites in the DFT+$U$ calculation. The values of $\Delta \epsilon_p$ are within $0.15$\,eV of the DFT calculated values.
The 2$p$ and 3$d$ spin-orbit splittings were calculated using a relativistic DFT+$U$ calculation. All parameters used here are compiled in Table \ref{tab:Slaterparameters}.

\begin{table*}[t]
    \caption{Summary of the double-counting correction, Slater-Condon integrals, and spin-orbit coupling parameters used
in the MLFT calculations (unit in eV). F$^0_{pd}$ is treated as a free parameter. An additional potential $\Delta \epsilon_p$ is introduced to shift the calculated $2p$ core energies to coincide with experimental values.}
    \begin{tabular}{l c c c c c c c c c c c c c c c l}
         & $\delta_{DC}$&& F$_{dd}^0$ & F$_{dd}^2$ & F$_{dd}^4$ && F$_{pd}^0$& F$_{pd}^2$ && G$_{pd}^1$ & G$_{pd}^3$ &&  $\zeta_p$ &$\zeta_d$ &&$\Delta \epsilon_p$\\\hline\hline
    CoFe$_2$O$_4$ Co        & 1.5   && 6.6 & 9.23 & 5.73 && 7.60  & 5.18 && 4.22 & 2.49 &&9.85 & 0.078 && 0\\
    CoFe$_2$O$_4$ Fe$_{\mathrm{oct}}$& 1.5      && 6.4 & 9.35 & 5.85 && 8.12 & 4.89 && 3.21 & 2.05 && 8.3 & 0.064 && 0\\
     CoFe$_2$O$_4$ Fe$_{\mathrm{tet}}$& 1.5     &&6.3 & 9.35  & 5.85 && 8.12 & 5.19 && 3.51 & 2.05 && 8.3 & 0.074 & &0.25\\
     NiFe$_2$O$_4$ Ni     & 1.5        && 6.9 & 10.53& 6.94 && 7.90  & 5.79 && 4.28 & 2.42 &&11.6 & 0.096 && 0\\
    NiFe$_2$O$_4$ Fe$_{\mathrm{oct}}$& 1.5      && 6.4 & 9.35 & 5.85 && 7.87 & 5.80 && 3.28 & 2.40 && 8.3 & 0.063 && 0\\
     NiFe$_2$O$_4$ Fe$_{\mathrm{tet}}$& 1.5      && 6.3 & 9.35 & 5.85 && 7.87 & 5.80 && 3.28 & 2.40 && 8.3 & 0.072 && 0.35\\ 
     Fe$_3$O$_4$ Fe$^{2+}_{\mathrm{oct}}$  & 1.5 && 6.4  & 8.63 & 5.36 && 8.45 & 4.69 && 3.18 & 1.92 && 8.3  & 0.062&& 4.02\\
      Fe$_3$O$_4$  Fe$^{3+}_{\mathrm{oct}}$ & 1.5 && 6.4  & 8.88 & 5.52 && 8.55 & 4.88 && 3.34 & 2.00 && 8.3  & 0.064 && 0\\
      Fe$_3$O$_4$ Fe$^{3+}_{\mathrm{tet}}$ & 1.5  && 6.3  & 9.28 & 6.26 && 7.82 & 5.21 && 3.57 & 2.14 && 8.3  & 0.076 && 1.35\\
      \hline\hline
    \end{tabular}
 
    \label{tab:Slaterparameters}
\end{table*}

\section{Crystal and magnetic structure}
\label{structure}
\begin{figure}
    \centering
    \includegraphics[scale=0.4]{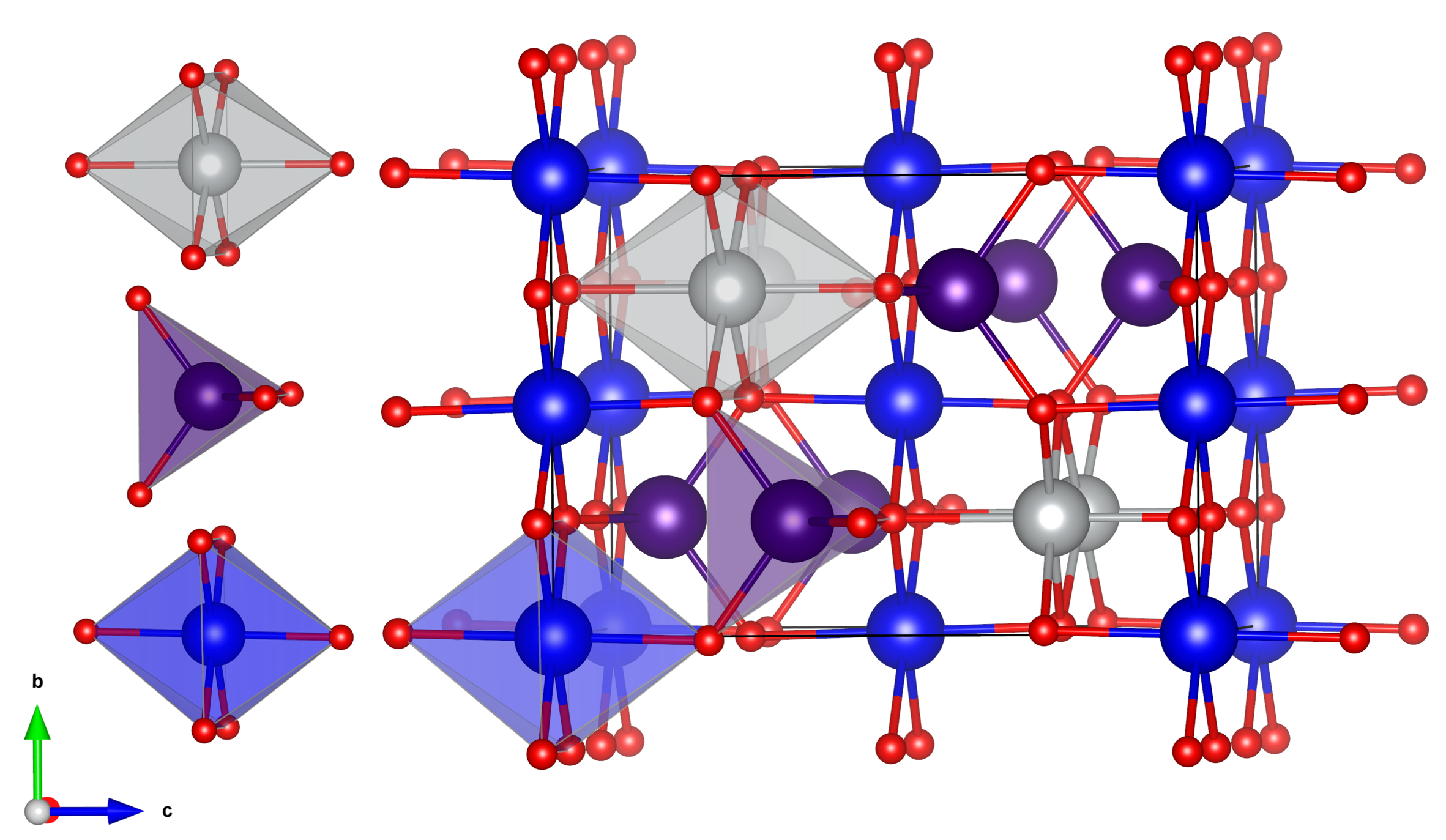}
    \caption{Conventional unit cell of XFe$_2$O$_4$, where X is Fe, Co or Ni. The silver-colored octahedral sites represent the Fe$^{2+}$ sites that are occupied by Co or Ni in the other ferrites
    . The blue octahedral sites and the purple tetragonal are occupied by Fe$^{3+}$.}
    \label{crystalstructure}
\end{figure}
Magnetite (Fe$_3$O$_4$) above the well known Verwey transition, has an inverse spinel structure, with 
an fcc Bravais lattice. 
For this structure, DFT+$U$ calculations, as presented here (for details, see Appendix \ref{appendixA}), result in a 
half-metallic electronic structure. The Fe cations are either octahedrally or tetrahedrally coordinated to the oxygen ions. Above the Verwey temperature, the six Fe ions in the unit cell are separated into two tetrahedral and four octahedral sites. The two tetrahedral sites and two of the octahedral sites are occupied by nominally Fe$^{3+}$ ions, while the remaining two octahedral sites are occupied by nominally Fe$^{2+}$ ions.
When we go from magnetite to CoFe$_2$O$_4$ or NiFe$_2$O$_4$, almost all Co/Ni atoms occupy the Fe atom at the octahedral +2 sites.
Therefore the structures can be expressed as (Fe$^{3+}$)[X$^{2+}$Fe$^{3+}$]O$_4$, where the X is Fe/Co/Ni and the parentheses indicate tetrahedral sites and the square brackets denote the octahedral sites. A schematic overview of the inverse spinel structure and the difference between the compounds
is shown in Fig.~\ref{crystalstructure}.
The Fe$_3$O$_4$ and the other ferrites are ferrimagnetic spinels. The magnetic ordering results from the exchange interactions between the cations. The Fe 3$d$ orbitals of the tetrahedral and octahedral sites overlap with the intermediate 2$p$ oxygen at an angle of $\approx$ 125$^\circ$, which according to the Goodenough-Kanamori-Anderson rules leads to antiferromagnetic exchange between the two sublattices. The octahedral sites couple ferromagnetically to each other due to a 90$^\circ$ superexchange. The exchange interaction between the tetrahedral sites is antiferromagnetic, however it does not result in an antiferromagnetic ordering in the tetrahedral sites as the other exchange interactions are stronger~\cite{Kim1999}. In Table \ref{moments-1} we compare the calculated spin and orbital moments with a previous calculation from Ref.~\cite{Antonov2003}. It can be seen that the two sets of calculations compare rather well with each other. 
In Table \ref{moments} the calculated sizes of the total magnetic moments of the tetrahedral sites and the average moment of the octahedral sites are compared to neutron scattering experiments~\cite{C8NR01534A} for Fe$_3$O$_4$, CoFe$_2$O$_4$ and NiFe$_2$O$_4$~\cite{PhysRevB.74.012410} taken at a temperature of 300 K. 
It is clear from the table  that the theoretical results are in good agreement with experiment. From Table \ref{moments} it is also clear that the results of magnetic moments from DFT and MLFT are in very good agreement with each other. Note that the values obtained from MLFT are obtained as expectation values of the spin and orbital angular momentum operators, using the ground-state wavefunction (without core hole excitation). The fact that the two sets of calculations give similar results is rewarding since it reflects on the accuracy on the mapping to the local Hamiltonian. 

\begin{table*}
\caption{DFT+$U$ calculated absolute value of the spin and orbital (in parenthesis) magnetic moments, in units of Bohr magneton ($\mu_B$) per atom compared with DFT+$U$ calculation of Ref.~\cite{Antonov2003}.}
\begin{tabular}{c c c c c c c c c c c c}
\centering
      &\multicolumn{3}{c} {CoFe$_2$O$_4$} && \multicolumn{3}{c} {NiFe$_2$O$_4$}&& \multicolumn{3}{c} {Fe$_3$O$_4$}\\
     Calculation &  Co & Fe$_{\mathrm{oct}}$& Fe$_{\mathrm{tet}}$ && Ni &Fe$_{\mathrm{oct}}$ & Fe$_{\mathrm{tet}}$&& Fe$^{2+}_{\mathrm{oct}}$&  Fe$^{3+}_{\mathrm{oct}}$ &  Fe$^{3+}_{\mathrm{tet}}$\\
      \hline
      \hline
Present & 2.53(0.00)             & 4.09(0.00)                    &  3.93(0.00) && 1.66 (0.34) & 4.07 (0.03) & 3.92 (0.03) &&3.59 (0.02) & 4.08 (0.02) & 3.82 (0.02)\\ 
Ref.~\cite{Antonov2003}& 2.57 (0.01) & 4.04 (0.04) & 3.90 (0.03) && 1.54 (0.27) & 4.09 (0.03) & 3.99 (0.02) &&3.54 (0.02) & 4.00 (0.02) &3.84 (0.02) \\
\hline\hline
\end{tabular}
\label{moments-1}
\end{table*}

\begin{table*}
\caption{Calculated DFT+$U$ total magnetic moments, in unit of Bohr magneton ($\mu_B$) per atom, of the different sublattices compared with the experimental results for CoFe$_2$O$_4$ and NiFe$_2$O$_4$ of Ref. \cite{C8NR01534A} and for Fe$_3$O$_4$ of Ref. \cite{PhysRevB.74.012410}. In the table $\mu_{\mathrm{oct}}$ is the average moment on the octahedral sites. }
\begin{tabular}{c c c c c c c c c}
\centering
 & \multicolumn{2}{c} {Fe$_3$O$_4$}&& \multicolumn{2}{c} {CoFe$_2$O$_4$} && \multicolumn{2}{c} {NiFe$_2$O$_4$}\\\hline \hline
        &   $\mu_{\mathrm{tet}}$ &  $\mu_{\mathrm{oct}}$ &&  $\mu_{\mathrm{tet}}$ &  $\mu_{\mathrm{oct}}$ &&  $\mu_{\mathrm{tet}}$ &  $\mu_{\mathrm{oct}}$ \\
\hline

DFT+$U$   &    4.23       & 3.85     &&    3.93       &   3.31      &&     3.95&   3.05     \\
MLFT ground state &   4.25     &  4.03    &&    4.23       &   3.11      &&     4.03     &  3.43      \\
Expt.     &  4.0        & 3.6      &&    3.5        &   3.24      &&     3.72     &    2.90    \\
\hline \hline
\end{tabular}
\label{moments}
\end{table*}

\section{Results and discussion}
\label{results}
In this section, 
the theoretical calculations of XAS and XMCD are compared with the experimental results (black curves), for all compounds of this investigation. The theoretical calculation is obtained by varying the higher-order Slater parameters ($k \geq 1$) in Eq.~(\ref{eq:Slaters}) individually between 70\% and 90\% of their calculated values while making the best fit to the experimental data. These parameters are shown in Table~\ref{tab:Slaterparameters} and the theoretical spectra based on them are labelled ''theory'' in Figs.~\ref{fig:Fe3O4}-~\ref{fig:NiCV}, and are shown as a red line.
In Figs.~\ref{fig:Fe3O4}-~\ref{fig:NiFeO-Ni} we also show results from an alternative theoretical calculation, where all the higher-order Slater parameters are screened to 80\% of their bare values (blue dashed line in Figs. ~\ref{fig:Fe3O4}-~\ref{fig:NiFeO-Ni}). Note also that in each subsection, we start with an analysis of XAS which is followed by the results of the XMCD. In the case of CoFe$_2$O$_4$ and NiFe$_2$O$_4$ the spectral features from the Fe sites will be presented first.\\

\noindent
\subsection{Spectra of Fe$_3$O$_4$}

\begin{figure}[htp]
\centering{
\begin{tabular}{c}
\includegraphics[scale=0.5]{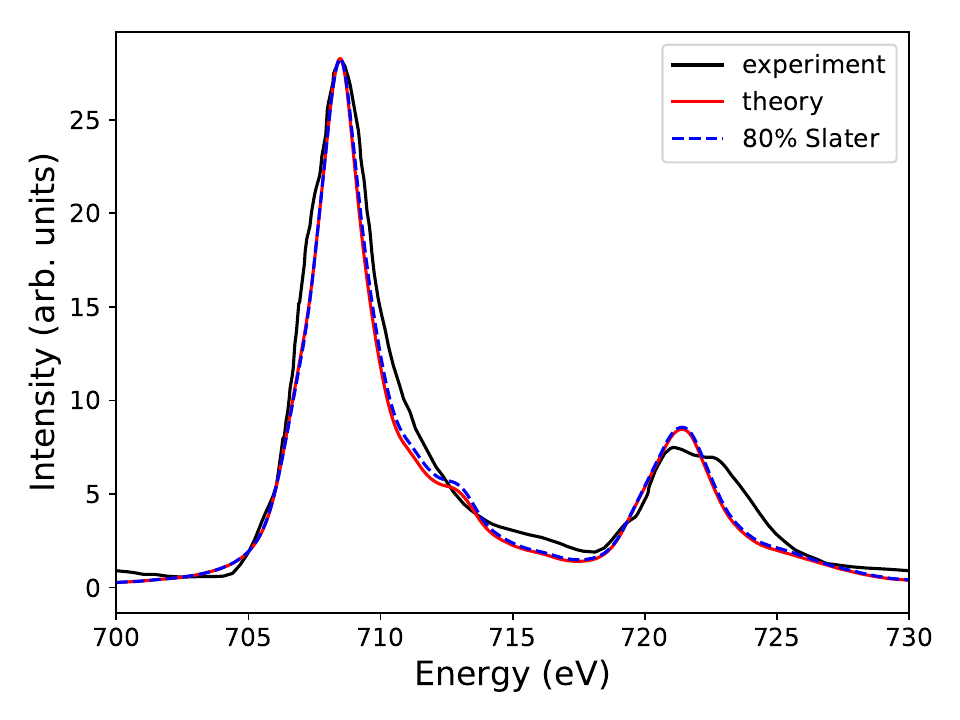}\\
\includegraphics[scale=0.5]{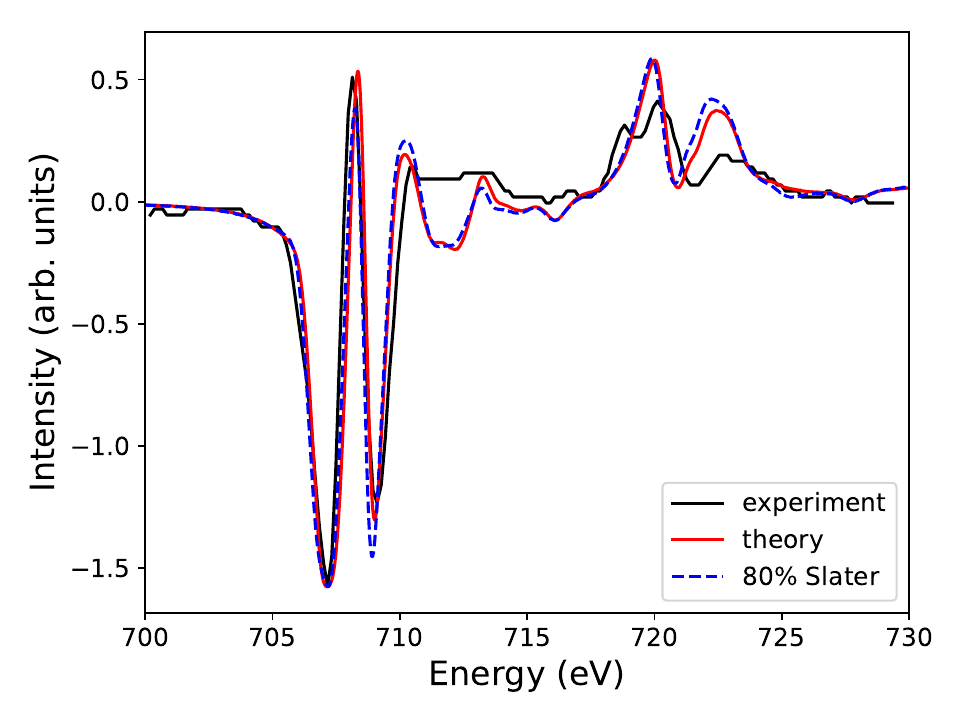}\\
\end{tabular}
}
\caption{Top: Calculated XAS of the Fe $L_{2,3}$ edges for Fe$_3$O$_4$ compared to experimental data (black curve). Bottom: Calculated XMCD of Fe$_3$O$_4$ compared to experimental data\cite{Richter2009} (black curve). The theory was carried out with the higher order Slater parameters screened between 70\%-90\% (red curve) and to 80\% (blue dashed curve) of the calculated values, for details see text.
}
\label{fig:Fe3O4}
\end{figure}

\begin{figure}[htp]
\centering{
\begin{tabular}{c}
\includegraphics[scale=0.5]{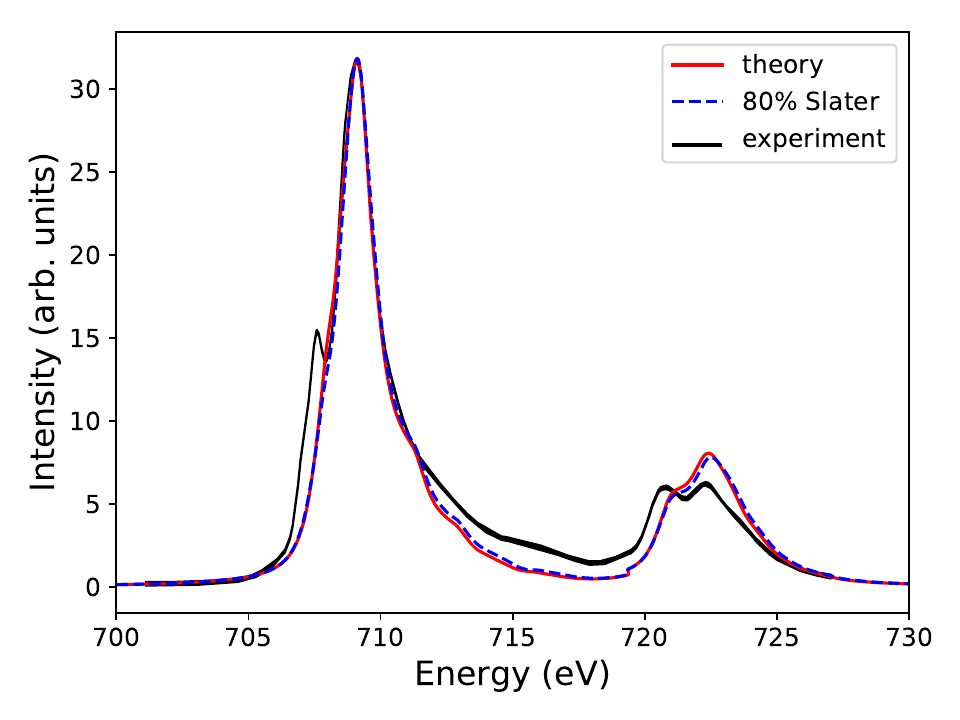}\\
\includegraphics[scale=0.5]{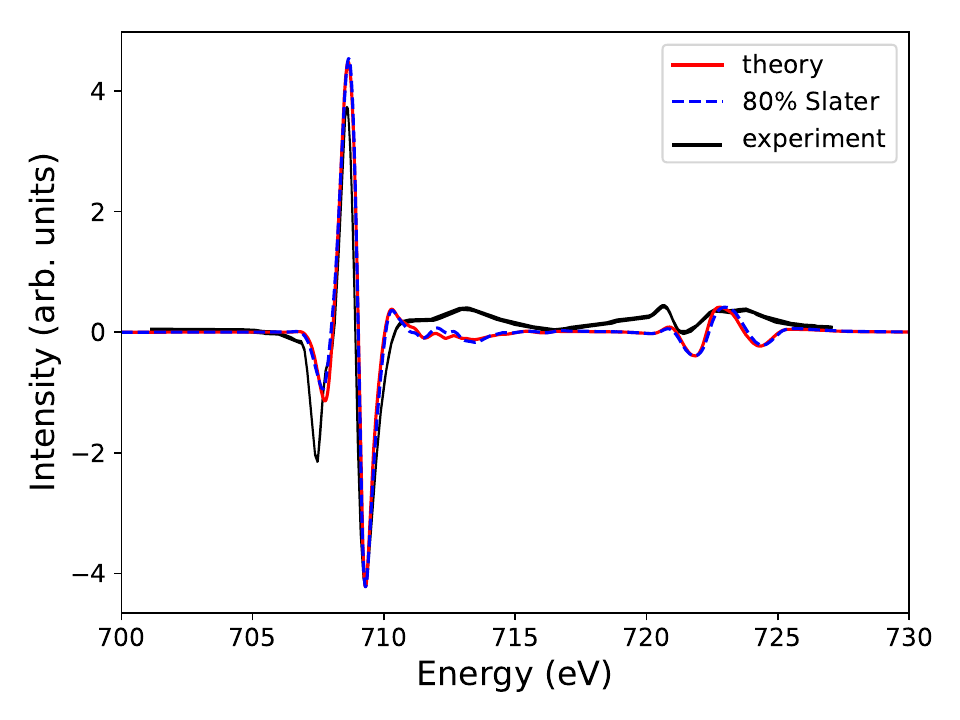}\\
\end{tabular}
}
\caption{Same as Fig.~\ref{fig:Fe3O4} but for Fe sites in CoFe$_2$O$_4$. The  experimental spectra are reproduced from~\cite{Wakabayashi2017}.
}
\label{fig:CoFeO-Fe}
\end{figure}

\begin{figure}[htp]
\centering{
\begin{tabular}{c}
\includegraphics[scale=0.5]{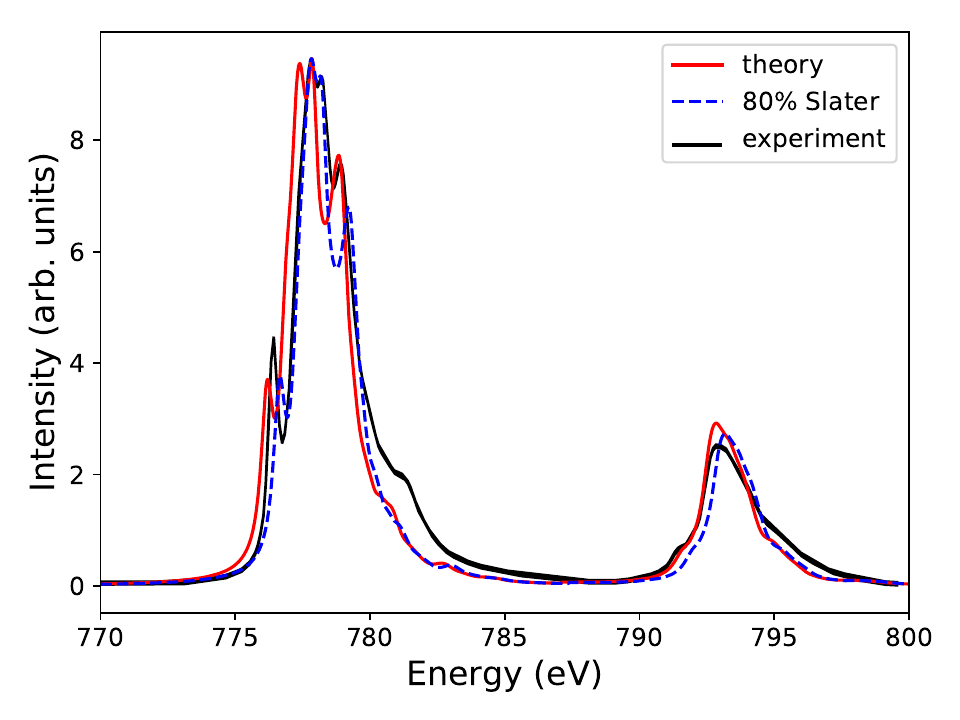}\\
\includegraphics[scale=0.5]{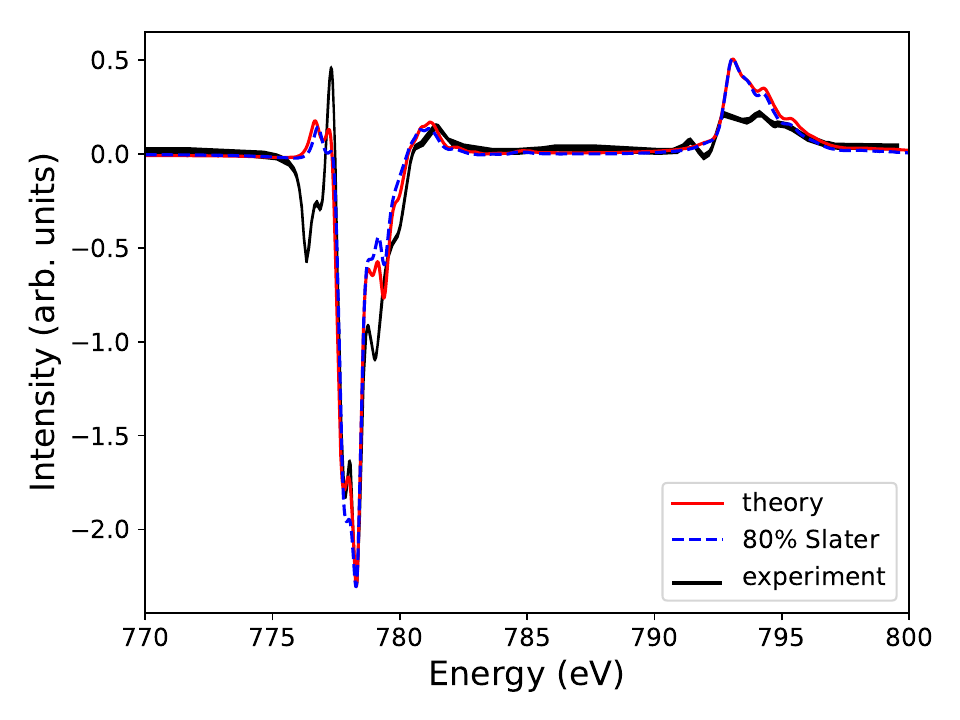}\\
\end{tabular}
}
\caption{Same as Fig.~\ref{fig:Fe3O4} but for Co site in CoFe$_2$O$_4$. The  experimental spectra is reproduced from~\cite{Wakabayashi2017}.
}
\label{fig:CoFeO-Co}
\end{figure}

\noindent
\textbf{XAS}.
The dipole allowed Fe $L_{2/3}$ XAS edges of Fe$_3$O$_4$ are displayed in Fig.~\ref{fig:Fe3O4}. The spectrum is composed of  two main peaks, the $L_3$- and $L_2$-edges, due to  the  SOC
$\Delta_{SO}=(3/2)\zeta_p$ in the TM 2$p$ state. This results in a split of the core-levels of 2$p$ $J$ =3/2 
and 2$p$ $J$ =1/2 by $\Delta_{SO}$. When photons excite
the core electrons in the higher  $J$ =3/2 level and lower $J$ =1/2 level they respectively produce 
electronic transitions to the $L_3$-edge  and the $L_2$-edge.
The SOC of the core states is strong and almost independent of the atomic environment.

The experimentally observed spectrum has a broad $L_{3}$-edge (between 705 and 715\,eV), which is composed of contributions from all three different Fe sites. 
The $L_2$-edge (between 718 and 728\,eV) has a significantly lower intensity and shows a double-peak feature.
Both edges are well reproduced by the theory with the $L_3$-edge being a bit broader on the low energy side in the experiment. The theoretical results from the $L_2$-edge do not show the same double-peak feature as in the experiment with the second peak missing, however, the relative intensity to the $L_3$-edge is more or less the same in theory and in the experiment. In a pure one-electron picture, and with uniform matrix elements for the excitation, the intensities of the $L_{3}$- and the $L_{2}$-edges would reflect the number of electrons available of the corresponding core levels, i.e., 4 to 2. The data in Fig.~\ref{fig:Fe3O4} show that this is not the case, and the so called branching ratio deviates significantly from the 4/2 ratio. It is gratifying that the theory used here reproduces this aspect of the measured spectrum. Furthermore, we can see that the calculation is robust towards the choice of Slater parameters as the curves match nearly perfectly.

\noindent
\textbf{XMCD}.
The measured XMCD spectrum (also shown in Fig.~\ref{fig:Fe3O4}) has a characteristic down-up-down peak structure at the $L_3$-edge.
Here, the first peak has the largest intensity and the second peak the lowest. This structure is caused by the ferrimagnetic ordering and the spinel structure. In a perfect antiferromagnet no XMCD can be observed as the contributions vanish. The antiferromangnetically aligned Fe cations occupy inequivalent lattice sites and have therefore slightly different excitation energies. This results in a dichroic response from the +3 cations, which here adds to the response from the Fe$^{2+}$ cations. At energies just above these peaks one can observe a small shoulder. The $L_2$-edge shows two different peaks with the first feature having a larger intensity and a distinct two-peak feature.
The theoretical calculation reproduces the behaviour at the $L_3$-edge very well. Even the small shoulder at energies above the three-peak feature is reproduced and the oscillations right below of the $L_2$-edge are reproduced. The only major difference is that a negative peak appears in the calculation between the $L_3$- and $L_2$-edge which is not in the experiment. Also the intensities of the $L_2$-edge is overestimated on all peaks except the first peak of the double-peak feature. This could be due to an overestimation of the 3$d$ SOC of the tetrahedral sites. Here, we can see that our choice of the higher order Slater parameters enhances the signal at the middle peak and suppresses that of the third peak and the multiplet effects.\\

\noindent
\subsection{Spectra of CoFe$_2$O$_4$}

\noindent
\textbf{Fe-projected results}\\

\noindent
\textbf{XAS}.
The measured XAS of the Fe sites (Fig.3, upper panel) shows a main peak with a low energy shoulder at the $L_3$-edge and a double peak at the $L_2$-edge. 
The XAS of the Fe sites shows the same features in experiment and theory. However, the shoulder of the $L_3$-edge is smaller in intensity for the calculated spectrum than in the experiment. This shoulder is caused by the octahedral sites in the calculation, therefore a higher ratio of octahedral to tetrahedral Fe sites than the equal amount that the calculation used could explain this effect. A reason for this could be Co cations occupying one or both tetrahedral sites in some cells, causing the Fe sites to occupy three or four octahedral sites in that cell, which would lead to a higher contribution of the octahedral Fe sites in the signal. Other point defects in the experimental sample could cause different oxidation states, that also would influence smaller details in the observed spectra, that would not be picked up by calculations of a defect-free system. For the $L_2$-edge the theory clearly puts more intensity on the higher energy peak, while they have the same intensity in the experiment. It is also notable that the branching ratio is very similar for theory and experiment. 
Furthermore, one can see that the calculation is robust to the choice of Slater parameters as the two curves match.\\

\noindent
\textbf{XMCD}.
In the XMCD spectrum (Fig.3, lower panel), one can observe the same characteristic down-up-down structure as for Fe$_3$O$_4$ in the $L_3$ channel, however due to the absence of Fe$^{2+}$ the first peak is no longer the largest in intensity, rather it is the smallest while the third peak in the $L_3$ channel has the largest intensity, with the second peak having similar intensity to the third. The details of these peaks are quite similar if one compares theory and experiment. The measured $L_2$-edge shows two low intensity peaks with the second one being flat and lower in intensity.
In the calculation the general structure of the $L_3$-edge is well reproduced. However, the intensity for the first peak is underestimated and the second one is overestimated, this could also be linked to some Co occupying tetrahedral sites, because the second peak is caused by tetrahedral Fe and the first by octahedral. Here, the fine structure is not reproduced by the calculation as it points in the opposite direction compared to the experiment. The two distinct $L_2$ peaks are well reproduced by the theory. However, the second peak in the experiment is broader than in the theory.
Furthermore, one can see that the calculation is robust to the choice of Slater parameters as the two curves match.\\

\noindent
\textbf{Co-projected results}\\

\noindent
\textbf{XAS}.
The XAS of Co (Fig.4, upper panel) shows a lot of features, both in theory and experiment. The $L_3$-edge has a double peak, where the peak on the left is slightly higher in intensity. The peak is accompanied by a small shoulder on the left, a high intensity shoulder and a small shoulder to the right. The $L_2$-edge is characterized  by a single peak.
The calculated XAS of the Co site agrees very well with the experiment. The double-peak feature of the $L_3$ is well reproduced, however in the experiment the first is larger. The left shoulder is located at lower energies in theory compared to  the experiment and it has also a lower intensity. The right shoulder matches the experiment perfectly, even the small shoulder at the end is reproduced.
At the $L_2$-edge, experiment and theory mostly lie on top of one another, however the theory shows a slightly higher intensity at the maximum of the peak.
Here, one can see that the two theoretical spectra look very similar and that the main difference is a shift of energy. This is caused by the fact that we are aligning the peaks of the largest intensity in the calculation to the largest peak in the experiment and the intensities of the two peaks in the double peak feature changes due to the choice of the Slater parameters.\\

\noindent
\textbf{XMCD}.
The measured $L_3$-edge XMCD of the Co site (Fig.4, lower panel) shows a small down-pointing peak followed by a thin up-pointing peak in front of the main peak, which has one shoulder on the left and two on the right.
After the main peak, there is a small fine structure peak. At the $L_2$-edge, there is a single flat peak.
The theory reproduces the shape of the $L_3$-edge main peak and its shoulders well, even though the shoulders on the right have less intensity than in the experiment. The peaks before the main peak are not well reproduced, with the first peak missing and the second peak being broader and less intense than in the experiment. This could be because these peaks are caused by Co occupying the tetrahedral site, where it would be closer to a +3 state. Alternatively, other impurities in the sample could play a role. The small fine structure after the main peak is very well resolved by the calculation. The $L_2$-edge in the theory shows a very clear peak with much higher intensity than the flat peak in the experiment. This could again be caused by Co in tetrahedral sites as those would have the opposite spin and therefore would reduce the signal from the peak.
Here, we can see that the choice of Slater parameters only affects the two shoulders closest to the main peak, where the intensity is shifted slightly from the left shoulder to the one on the right.\\

\noindent
\subsection{Spectra of NiFe$_2$O$_4$}

Here, the experiment uses a superposition of 77\% circular polarized and 23\% linear polarized light~\cite{Knut2021}. This was explicitly taken into account in the transition dipole operator. \\

\noindent
\textbf{Fe-projected results}\\

\noindent
\textbf{XAS}.
The measured $L_3$-edge of the Fe sites (Fig.5, upper panel) has a main peak with a shoulder to the left and a smaller shoulder to the right.
The $L_2$-edge has a clear main peak with a distinct shoulder to the left.
In the calculations of the XAS of the Fe sites the left shoulder of the $L_3$-edge is not present. However, the right shoulder is nearly perfectly reproduced by the theory. The observed relative intensity of the $L_2$-edge compared to the $L_3$-edge is slightly overestimated by our theory and the calculated shoulder of the $L_2$-edge could be more distinct from the main peak. Furthermore, the calculated energy difference between the $L_3$- and $L_2$-edges, due to the SOC of the $p$-electrons, is slightly overestimated.
Here, the blue and red lines perfectly match as we have used the same parameters.\\

\noindent
\textbf{XMCD}.
The measured XMCD (Fig.5, lower panel) shows the same down-up-down structure as the other Fe sites. However, here, the up-pointing peak has the highest intensity and the first peak has the lowest due to the missing Fe$^{2+}$ sites. After the three peaks, one can see two fine structure features. The $L_2$-edge is characterized by an up-down-up structure with the down-pointing peak having the highest intensity and the last peak the lowest and most elongated.
The down-up-down peak structure of the $L_3$-edge is reproduced very well by the theory with a slight shift towards higher energies for the first two peaks and an overestimation of the middle peak. The two fine-structure peaks after the third peak are also reproduced, however the intensity of the first peak is overestimated and of the second underestimated. In the $L_2$ channel, one can notice the same structure in the first two peaks from the calculations, but for the third we see a double-peak structure with higher intensities. Just like for the XAS, the SOC is overestimated and therefore, the energy difference between the $L_3$- and $L_2$-edges are slightly larger than in the experiment.\\

\noindent
\textbf{Ni-projected results}\\

\noindent
\textbf{XAS}.
The observed $L_3$-edge of Ni (Fig.6, upper panel) is characterized by a main peak with a large shoulder on the high energy side. The $L_2$-edge has a peak with a shoulder on the high energy side.
Overall, the XAS is well reproduced by the theory.
The main peak and shoulder of the $L_3$-edge are located at the correct positions, however the width and intensity are somewhat underestimated.
The shape and relative intensity of the $L_2$-edge compared to the $L_3$-edge are very close to the experimental results. However, the intensity is slightly lower than in the experiment.
Additionally the energy difference between the $L_3$ and $L_2$ peaks, which is determined by the SOC of the $p$-states, is slightly overestimated by the theory.
From Fig.6 one can note that the calculated result is robust with respect to an uncertainty in Slater parameters, as the intensity of the shoulders increases only slightly when the paramaters are varied. 

To further check the sensitivity of the theoretical results, we performed additional calculations that demonstrate how the theoretical spectra depend on the Coulomb core-valence (CV) interaction. This is shown in Figs. \ref{fig:FeCV} and \ref{fig:NiCV}, for the Fe respectively Ni sites in NiFe$_2$O$_4$. As the figure shows, 
it is crucial to include the CV interaction in order to get a good XAS and XMCD signal and the effect is most clearly demonstrated for the Fe atom. \\

\noindent
\textbf{XMCD}.
The XMCD (Fig.6, lower panel) has a large, up-pointing shoulder at the $L_3$-edge. At the $L_2$-edge, we observe a large peak with a shoulder to the left.
The calculation nearly perfectly matches the experimental results. The shoulder of the $L_3$-edge is slightly too low in intensity, whereas the energy difference between the $L_3$- and $L_2$-edges is slightly overestimated like in the XAS. Here, we can see that the calculation is just as robust to the choice of Slater parameters as the XAS.

To summarize the previous discussion,  we observe that all  XAS excitations
can be explained in terms of  electric-dipole-allowed electronic transitions 
and  are composed of  two strong peaks in  the $L_3$ and $L_2$
edges, due to the large SOC of  the 2$p$ states into $p_{3/2}$
and $p_{1/2}$ relativistic states.  The photon excitations of the core electrons in
the higher  $J$ =3/2  and low $J=1/2$  levels produce electronic transitions
essentially towards the unoccupied 3$d$ states, producing respectively the
$L_3$-   and $L_2$-edges. The features  within each of those  edges are
mainly   due to the 3$d$ electron-electron  interactions as well as
the Coulomb interaction between   the 2$p$ core hole and the  photo-electron in
the 3$d$ unoccupied states. It is therefore a good idea to decompose such
excitations in terms of the   3$d$ $t_{2g}$ and $e_g$ states to determine the
various contributions to the XAS and XMCD spectra. For this we have inspected the partial density of states (PDOS) to investigate which orbitals resemble $e_g$ and $t_{2g}$ the most.
Then we constructed operators that only excite the 2$p$ core electrons into either the $e_g$ or $t_{2g}$ orbitals. 
The difference between the total spectra and the sum of the signals from $e_g$ and $t_{2g}$, corresponds to the interference between the excitations into $ t_{2g}$  and $e_g$ that produce the same final state. We refer to this as the off-diagonal contribution in the following.

Figure \ref{fig:Fe3O4-egt2g}
 shows for Fe$_3$O$_4$
the contributions from excitations between $2p_{1/2}$ as well as the $2p_{3/2}$ core levels to the $ t_{2g}$  and $e_g$
states, resulting in symmetry projections of the final states reached by the light matter interaction. Note from Fig.~\ref{fig:Fe3O4-egt2g} that we show both data for XAS as well as XMCD.   Similar figures for the analysis of the XAS and XMCD spectra of 
CoFe$_2$O$_4$ and NiFe$_2$O$_4$
are shown in the supplementary information. It is clear from Fig.~\ref{fig:Fe3O4-egt2g} that
the intensity of the two symmetry components is not quite a 3/2 ratio  that would be expected from the occupancy of the  $t_{2g}$/$e_g$ levels, but
only about 1.4. One may also see that the $t_{2g}$  and $e_g$ energy splitting is not noticeably conspicuous in these spectra. It is
interesting to note that both states contribute almost equally to the $L_3$
and $L_2$ main peaks. For the XMCD spectrum, the difference between $t_{2g}$ and $e_g$ projected levels is much more pronounced compared to the XAS signal. In particular we note that for the $L_2$-edge the $ t_{2g}$ and $e_g$ have opposite contributions at most of the energies where this peak has large intensity. In general, however, Fig.~\ref{fig:Fe3O4-egt2g} shows that an interpretation of the XAS and XMCD signals 
in terms of transitions to $t_{2g}$  or $e_g$ projections is not trivial, since the off-diagonal contributions are significant. Similar conclusions can be drawn for symmetry decomposed spectra of all the here investigated materials (see Appendix~\ref{sec:greygoose}). 

As a final comment to this section we note that the XAS and XMCD spectra of NiFe$_2$O$_4$ are very similar for calculations where the parameters of the MLFT Hamiltonian are evaluated from LDA$+U$ and from LDA+DMFT. This is shown in Fig.14 for the Fe L-ege spectra and in Fig.15 for the Ni spectra. Note that the two levels of theory give almost identical results for the Fe XAS signal, but that the XMCD signal agrees sightly better with experiment for the DMFT based theory. For the Ni signal the two theory curves are almost identical both for the XAS and XMCD signal.

\begin{table*}[t]
\caption{Site electron occupation $N_{3d}$ of the tetrahedral and the octahedral transition metal ion in different ferrites.}
\begin{tabular}{c c c c c c c c c c c c}
\centering
     &  \multicolumn{3}{c} {Fe$_3$O$_4$} && \multicolumn{3}{c} {CoFe$_2$O$_4$}&& \multicolumn{3}{c} {NiFe$_2$O$_4$}\\
      &  Fe$^{2+}_{\mathrm{oct}}$ & Fe$^{3+}_{\mathrm{oct}}$& Fe$^{3+}_{\mathrm{tet}}$ && Co &Fe$_{\mathrm{oct}}$ & Fe$_{\mathrm{tet}}$&& Ni & Fe$_{\mathrm{oct}}$ &  Fe$_{\mathrm{tet}}$\\
      \hline
      \hline
N$_{3d}$& 6.267         & 5.901            &  5.768         && 7.129 & 5.148 & 5.110 && 8.291 & 5.177 & 5.123\\ 
\hline\hline
\end{tabular}
\label{electrons}
\end{table*}


\begin{figure}[htp]
\centering{
\begin{tabular}{c}
\includegraphics[scale=0.5]{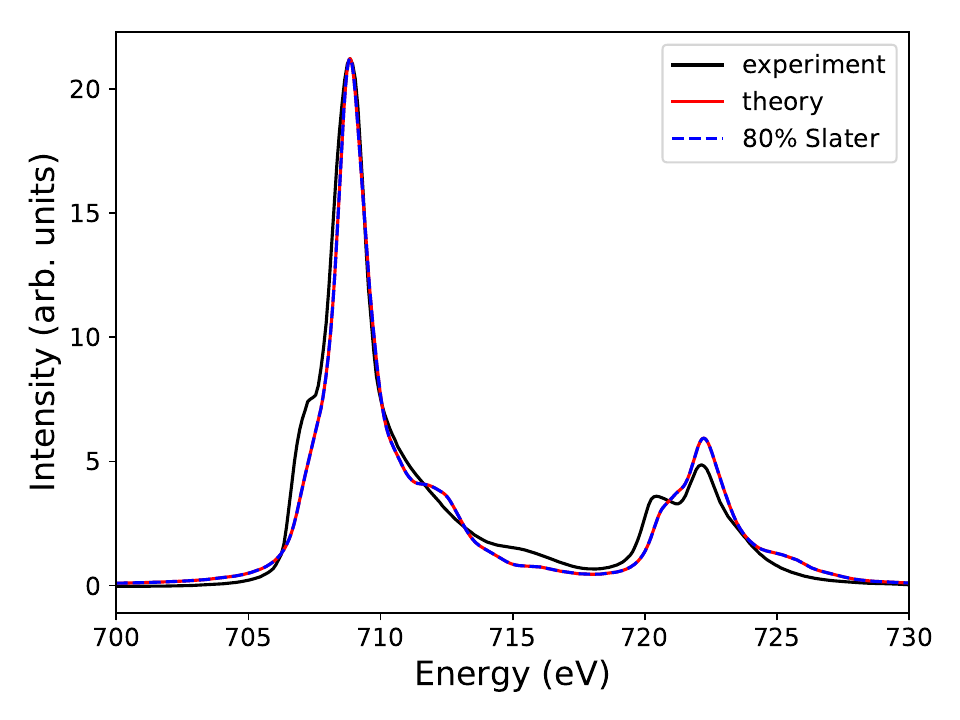}\\
\includegraphics[scale=0.5]{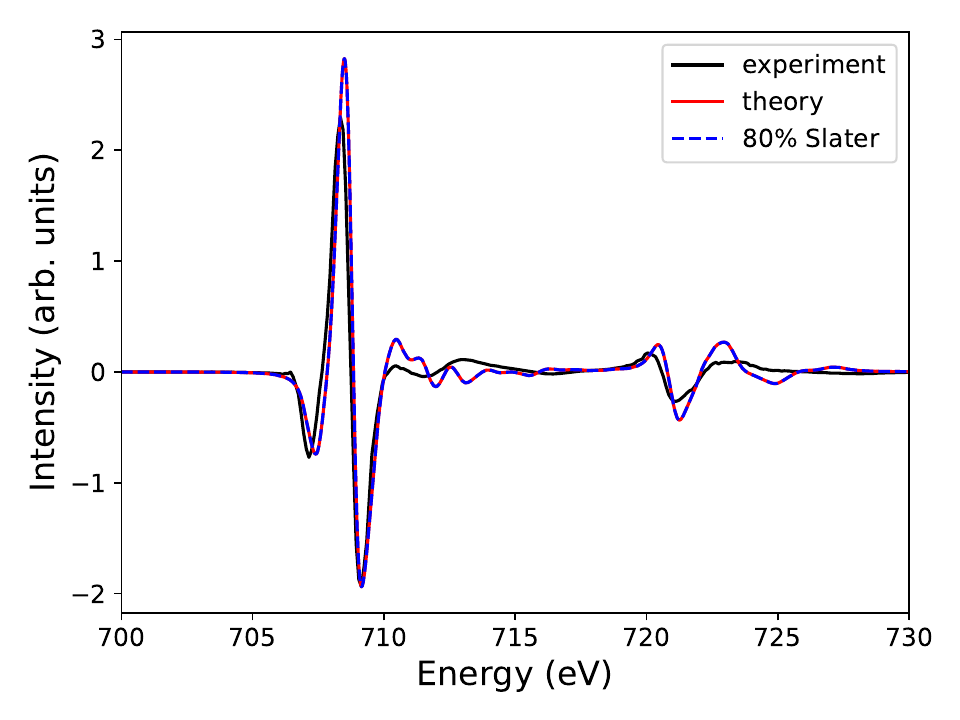}\
\end{tabular}
}
\caption{Same as Fig.~\ref{fig:Fe3O4} but for Fe sites in NiFe$_2$O$_4$. The  experimental spectra is reproduced from~\cite{DarioArenaexp}. 
}
\label{fig:NiFeO-Fe}
\end{figure}
\begin{figure}[htp]
\centering{
\begin{tabular}{c}
\includegraphics[scale=0.5]{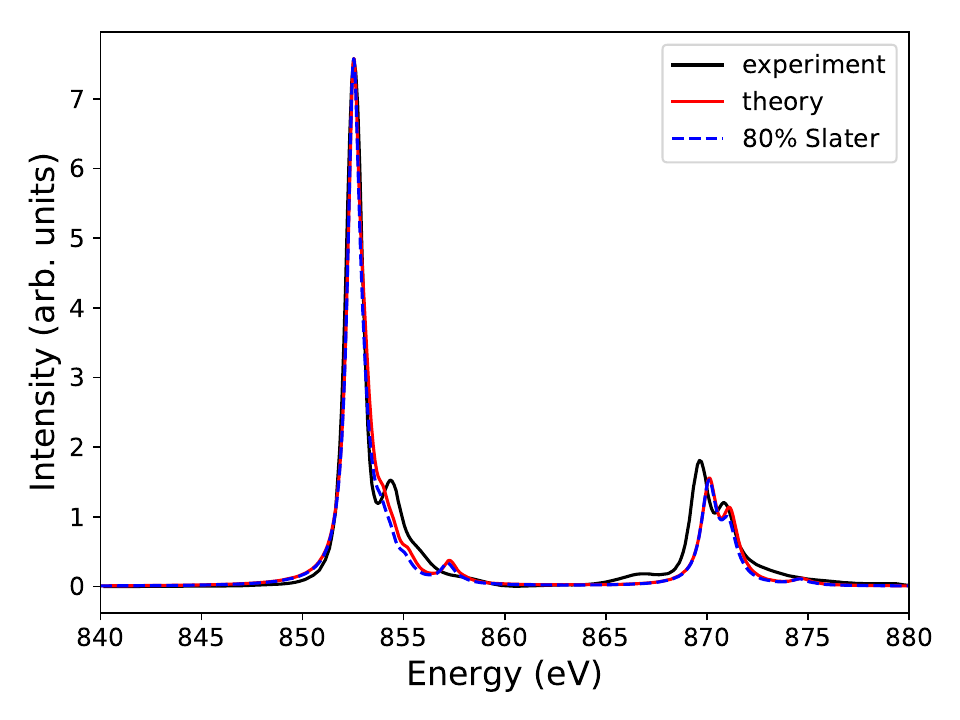}\\
\includegraphics[scale=0.5]{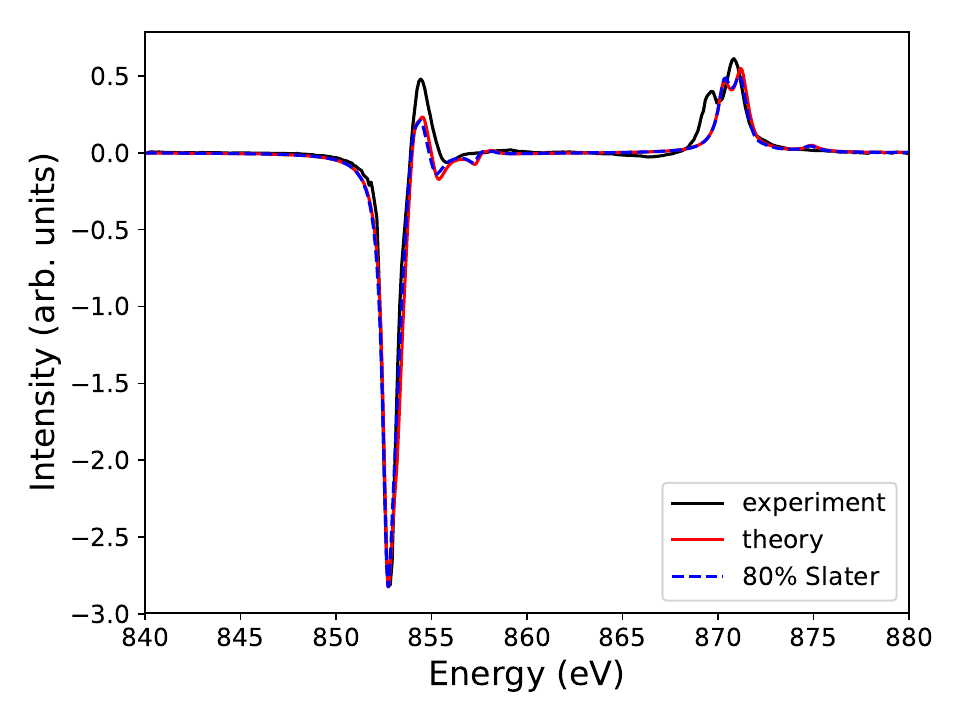}\\
\end{tabular}
}
\caption{Same as Fig.~\ref{fig:Fe3O4} but for Ni site in NiFe$_2$O$_4$. The  experimental spectra is reproduced from~\cite{DarioArenaexp}. 
}
\label{fig:NiFeO-Ni}
\end{figure}
\section{Conclusion}
We have used multiplet ligand-field theory in conjunction with DFT+$U$ or DFT+DMFT electronic structure theory to describe on an equal footing the multiconfigurational 
effects as well as the presence of the core-hole photo-electron Coulomb interaction and its effect on
XAS and XMCD spectra. 
The electronic  transition selection rules have been included within
the electric-dipole approximation. We have presented the details of the implementation
and its application to the  $L_{2,3}$-edges and XMCD spectra of the ferrites, Fe$_3$O$_4$ (magnetite), cobalt ferrite, CoFe$_2$O$_4$,
and nickel ferrite, NiFe$_2$O$_4$. We find that the results are in overall agreement  
 with available experimental spectra. In particular, we find that
the experimental $L_3$/$L_2$ branching ratio of the XAS signal is well reproduced by the here described
calculations, in contrast to an approximate 2:1 ratio produced by DFT calculations. In addition, most of the details of the spectral features of the measured XAS and XMCD signals are reproduced by the theory.
In addition, and as detailed in Appendix~\ref{appendix:DMFT}, our analysis suggests that calculations with parameters obtained from DFT+$U$ and DFT+DMFT give very similar results both for the XAS and XMCD spectra. If this can be demonstrated to hold for a wider range of compounds, it would in general simplify the theoretical description of XAS and XMCD spectra of materials with complex electronic structure, since the computational effort of DFT+$U$ is significantly smaller than DFT+DMFT.

The aim of the present theory is to rely on as few free parameters as possible as detailed in the method section~\ref{details}. 
We explored the renormalization of the higher-order Slater parameters due to screening by either renormalizing their values to a fixed 80\%  of their bare value or varying them between 70\% to 90\%. We found that the spectra were only marginally affected by different descriptions of the screening within this range. In this work, only the  zeroth-order Slater parameter $F_{pd}^0$ and the double-counting correction $\delta_{DC}$ were chosen as free parameters. We kept the value of $\delta_{DC}$ fixed to 1.5\,eV for all investigated compounds. There is currently no accurate method for calculating the screening of $F_{pd}^0$, so we varied it in the interval $F_{dd}^0< F_{pd}^0<1.4 F_{dd}^{0}$. We are currently investigating how to obtain screening of $F_{pd}^0$ from cDFT in a reliable way. 

We have attempted to make an interpretation in terms of excitations from  $2p_{1/2}$ and $2p_{3/2}$ to the $ t_{2g}$  and $e_g$ irreducible representations of the 3d valence electron 
states. However, due to large inteference terms between the $e_g$ and $t_{2g}$ excitations, such an analysis was shown to have a convoluted physical interpretation. 
\acknowledgements
We acknowledge support from the Swedish Research Council, Wallenberg Initiative Materials Science (WISE) funded by the Knut and Alice Wallenberg Foundation, the ERC (FASTCORR project), eSSENCE and STandUPP. Calculations provided by SNIC/NAISS. We are grateful to Dario Arena, University of South Florida, for allowing us to show his experimental data of NiFe$_2$O$_4$, prior to publication. Valuable discussions with Dr. Chin Shen Ong, Dr. Yaroslav Kvashnin and Prof. Mauritz Haverkort are acknowledged.

\appendix

\section{Computational details}
\label{appendixA}
The XAS and XMCD spectra were calculated for NiFe$_2$O$_4$, CoFe$_2$O$_4$ and Fe$_3$O$_4$, which all share the same inverse-spinel structure with Ni and Co occupying the octahedral sites.

\subsection{DFT}

Self-consistent  spin-polarized DFT+$U$ calculations, where the spin polarization was introduced via the local Hartree-Fock potential (+$U$), have been performed with $8\times 8\times 8$ $k$-point mesh sampling in the Brillouin zone for NiFe$_2$O$_4$ and CoFe$_2$O$_4$ and $10 \times 10 \times 10$ for Fe$_3$O$_4$. 
We used a full-potential linear muffin-tin orbital method (LMTO) as implemented in the ``RSPt" code\cite{rspt-book,rspt-web} to solve the DFT problem.
The set of localized impurity orbitals is constructed by projecting the total electron density on a set of L{\"o}wdin orthogonalized LMTOs for the TM $3d$ orbitals, denoted as ``ORT" in Ref.~\cite{PhysRevB.76.035107,PhysRevLett.109.186401,oscargranas}.

\subsection{Coulomb $U$}
The interacting part of the impurity Hamiltonian for a multi-orbital system reads
\begin{equation}
\label{eq:DMFT:Coulomb}
\hat{H}_\textrm{int.} =  \frac{1}{2} \sum_{abcd,\sigma\sigma'} U_{abcd} \hat{c}_{b,\sigma}^\dagger \hat{c}_{a,\sigma'}^\dagger \hat{c}_{c,\sigma'} \hat{c}_{d,\sigma},
\end{equation}
where each term describes a process and $U_{abcd}$ is given by
\begin{align}
\label{eq:Uijkl}
U_{ijkl} &=  \int \int d^3r d^3r'  \psi_i^*(\vect{r}') \psi_j^*(\vect{r}) \frac{1}{|\vect{r}-\vect{r}'|} \psi_k(\vect{r}') \psi_l(\vect{r}).
\end{align}

By expanding the Coulomb interaction $1/|\vect{r}-\vect{r}'|$ in terms of spherical harmonics and with basic functions of the form 
\begin{equation}
\label{eq:local:basis}
\psi_i(\vect{r}) = f_{n_i,l_i}(r) Y_{l_i,m_i}(\theta,\phi),
\end{equation}
the Coulomb interaction tensor becomes~\cite{Eder2012}
\begin{align}
\label{eq:CoulombExpansion}
U_{abcd} =\delta_{m_a+m_b,m_c+m_d} \sum_{k=0}^{k_\textrm{max}}  & c^k(l_b,m_b; l_d,m_d) c^k(l_c, m_c; l_a,m_a)  \nonumber \\  
& \times R^k(n_a l_a, n_b l_b, n_c l_c, n_d l_d). 
\end{align}
  
The Gaunt coefficients, 
\begin{align}
\label{eq:Gaunt}
c^k(l,m;l',m') = \sqrt{\frac{4\pi}{2k+1}} &\int_0^{2\pi}  d\phi \int_0^{\pi}  d\theta \sin{\theta} Y^*_{l,m}(\theta,\phi)    \nonumber \\
& \times  Y_{k,m-m'}(\theta,\phi) Y_{l',m'}(\theta,\phi), 
\end{align}
take care of the angular integrals in Eq.~\eqref{eq:Uijkl} and are easily evaluated. 
By considering the parity of the spherical harmonics in Eq.~\eqref{eq:Gaunt}, only Gaunt coefficients with $l+l'+k$ being an even number can be non-zero. 
The two Gaunt coefficients in Eq.~\eqref{eq:CoulombExpansion} constrain the $k$-expansion to a maximum of $k_\textrm{max} = \textrm{min}(| l_b+l_d |, |l_c+l_a|)$. 
The last factor to discuss in Eq.~\eqref{eq:CoulombExpansion} is the parameter,
\begin{align}
\label{eq:Slaterint}
R^k(n_a l_a, n_b l_b, n_c l_c, n_d l_d)& =   \int_0^\infty  \int_0^\infty  dr dr'   r^2 r'^2 f_{n_a,l_a}(r')   \nonumber \\  
& \times f_{n_b,l_b}(r) \frac{r_<^k}{r_>^{k+1}} f_{n_c,l_c}(r') f_{n_d,l_d}(r),
\end{align}
where $r_<$ ($r_>$) indicates min$(r,r')$ (max$(r,r')$).
It is customary to define the Slater-Condon parameters,  
\begin{align}
\label{eq:Slaters}
F^k(nl,n'l') &= R^k(nl,n'l',nl,n'l') \nonumber \\
G^k(nl,n'l') &= R^k(nl,n'l',n'l',nl), 
\end{align}
where $F$ and $G$ describe the Coulomb and exchange integrals, respectively. 
For the Coulomb interaction between $d$-orbitals, for any given principal quantum number, $F^k=G^k$, and only the three parameters $F^0, F^2$ and $F^4$ are relevant, due to the constrains mentioned above. The bare Slater-Condon integrals are calculated using the projected $3d$ and $2p$ wave functions within the muffin-tin sphere. However, the screened value of $F_{pd}^0$ is difficult to calculate due to the strong screening effects from uncorrelated electrons and is treated as a tunable parameter. \\

\begin{figure}
    \centering
    \includegraphics[scale=0.5]{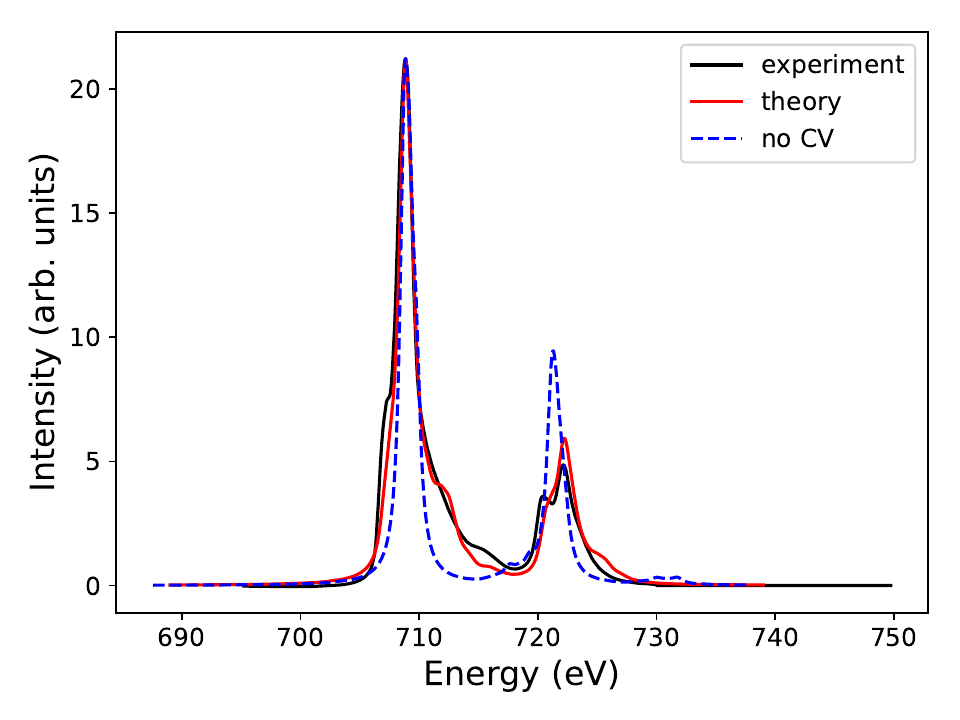}\\
    \includegraphics[scale=0.5]{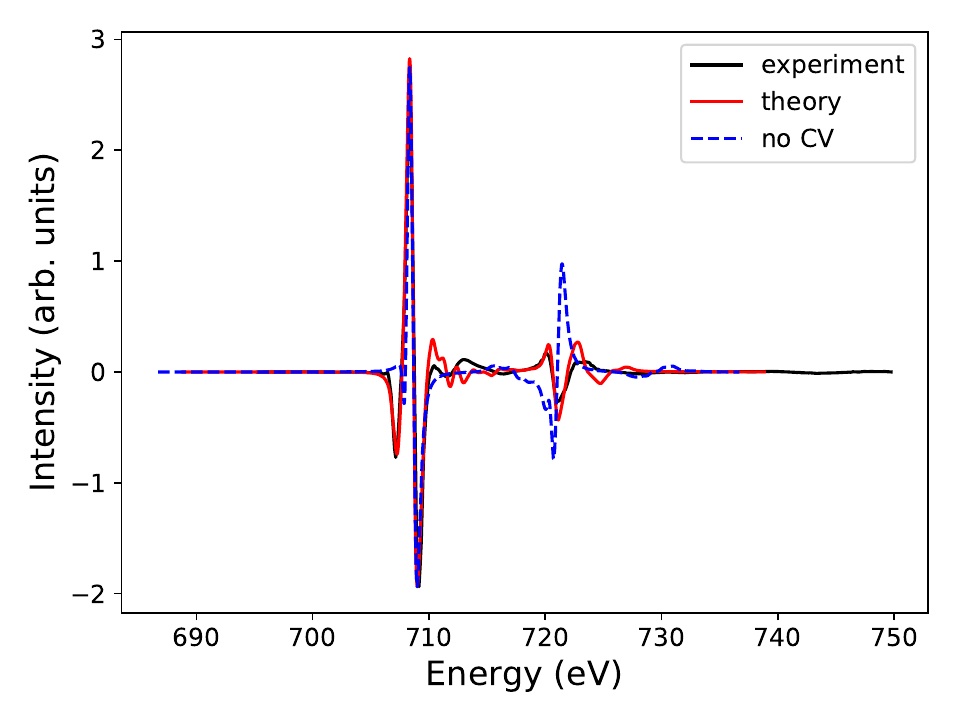}\\
    \caption{Calculated Fe $L_{2,3}$  XAS (top panel) and XMCD (bottom panel) edges in NiFe$_2$O$_4$ using the full Coulomb interaction (red solid) and neglecting the $p-d$ interaction (no CV, blue dashed) compared to experiments (black solid).}
    \label{fig:FeCV}
\end{figure}

\begin{figure}
    \centering
    \includegraphics[scale=0.5]{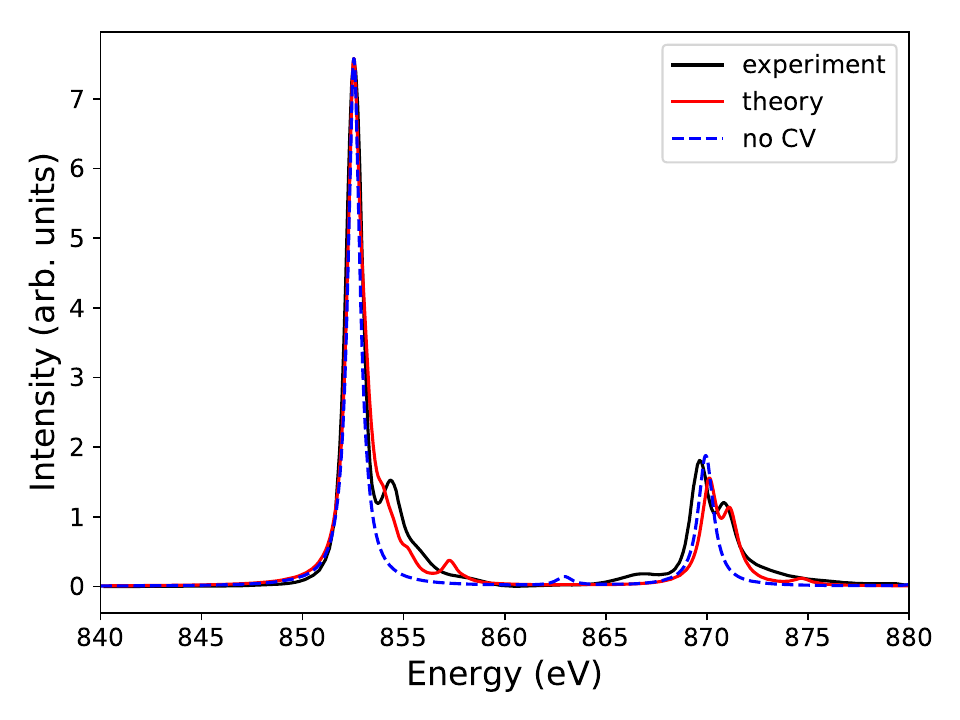}\\
    \includegraphics[scale=0.5]{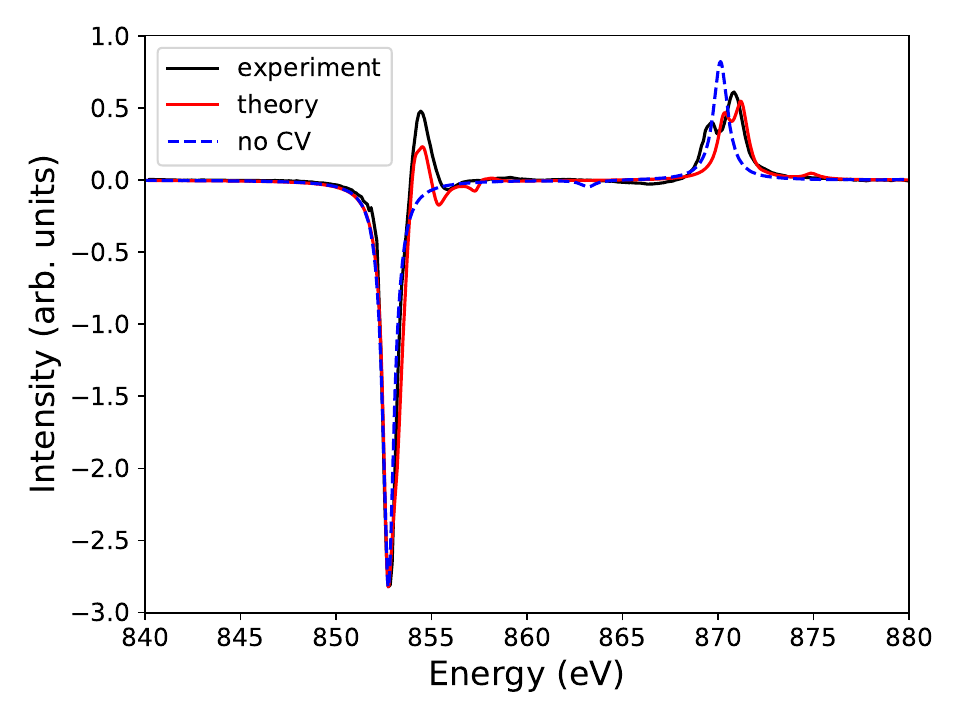}\\
    \caption{Calculated Ni $L_{2,3}$  XAS (top panel) and XMCD (bottom panel) edges in NiFe$_2$O$_4$ using the full Coulomb interaction (red solid) and neglecting the $p-d$ interaction (no CV, blue dashed) compared to experiments (black, solid).}
    \label{fig:NiCV}
\end{figure}

The core-valence interaction ($F^k_{pd}$ and $G^k_{pd}$) gives crucial contributions to the spectra. In Fig.~\ref{fig:FeCV} and Fig.~\ref{fig:NiCV} we compare the calculated spectra of NiFe$_2$O$_4$ using the full Hamiltonian and neglecting the core-valence interaction. The L$_{2,3}$-edges in the spectra without core-valence interaction are characterized by having a single Lorenzian peak, except the XMCD of Fe where we have a single peak per site.

\subsection{Spin-orbit coupling}
The spin-orbit coupling (SOC) Hamiltonian is first quantized form for a $(n,l)$-shell with $N$ electrons and SOC parameter $\zeta$ is  
\begin{equation}
\label{eq:XAS:SOC}
\hatF{H}_\textrm{SOC} = \zeta \sum_{i=1}^N  \hatF{\vect{l}}_i \cdot \hatF{\vect{s}}_i = \zeta \sum_{i=1}^N  \left ( \hatF{l}_i^z \hatF{s}_i^z + \frac{1}{2}(\hatF{l}_i^+ \hatF{s}_i^- + \hatF{l}_i^- \hatF{s}_i^+) \right ),
\end{equation} 
where for particle $i$, $\hatF{\vect{l}}_i$ ($\hatF{\vect{s}}_i$) is the orbital (spin) angular momentum vector operator, $\hatF{l}_i^z$ ($\hatF{s}_i^z$) the $z$-projected orbital (spin) angular momentum operator and $\hatF{l}_i^{\pm}$ ($\hatF{s}_i^{\pm}$) the raising and lowering orbital (spin) angular momentum operators.
In the last expression in Eq.~\eqref{eq:XAS:SOC}, the first term is diagonal in the $(l,m,\sigma)$ basis and in second quantized form becomes
\begin{equation}
\zeta \sum_{m=-l}^l \sum_{\sigma \in \{ -\frac{1}{2},\frac{1}{2}  \} }  \sigma m \hat{c}_{l,m,\sigma}^\dagger \hat{c}_{l,m,\sigma}.
\end{equation} 
The other terms flip the spin and can be written as
\begin{equation}
 \zeta  \frac{1}{2} \sum_{m=-l}^{l-1} \sqrt{(l-m)(l+m+1)} (\hat{c}_{l,m+1,\downarrow}^\dagger \hat{c}_{l,m,\uparrow} + \hat{c}_{l,m,\uparrow}^\dagger \hat{c}_{l,m+1,\downarrow}).
\end{equation} 
For $3d$ orbitals of the 3$d$ elements, $\zeta$ is rather small (less than $\sim$ 100\,meV) in comparison to other relevant energies, e.g., the bandwidth. But for core $2p$ orbitals of the 3$d$ transition metals $\zeta$ is of the order of several eV and thus absolutely necessary to include in the calculation.

\subsection{Double counting \label{appendix:DC}}

Another important aspect is the double counting (DC) correction $\delta_{DC}$, which has to be subtracted from the DFT-derived Hamiltonian. This is done in order to remove the contribution of the Coulomb repulsion that is already taken into account at the DFT level. 

In this work, we apply a DC that is often used in MLFT by considering the relative energy for different configurations~\cite{PhysRevB.96.245131}. 

\section{Supporting Results}
\subsection{Symmetry-decomposed spectra}
\label{sec:greygoose}

\begin{figure}[htp]
\centering{
\begin{tabular}{c}
\includegraphics[scale=0.5]{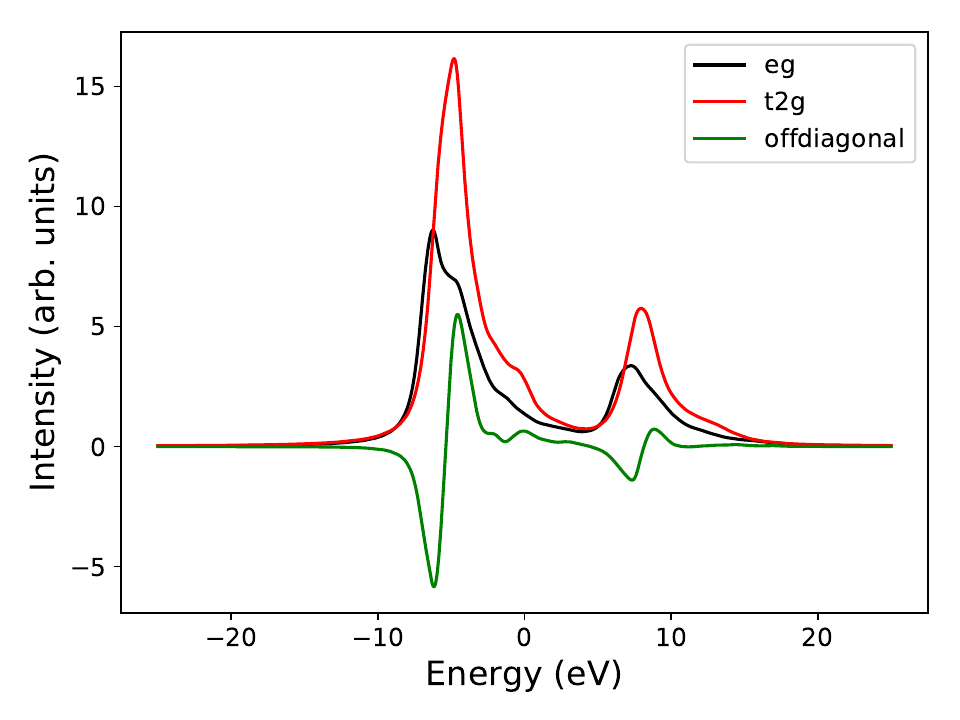}\\
\includegraphics[scale=0.5]{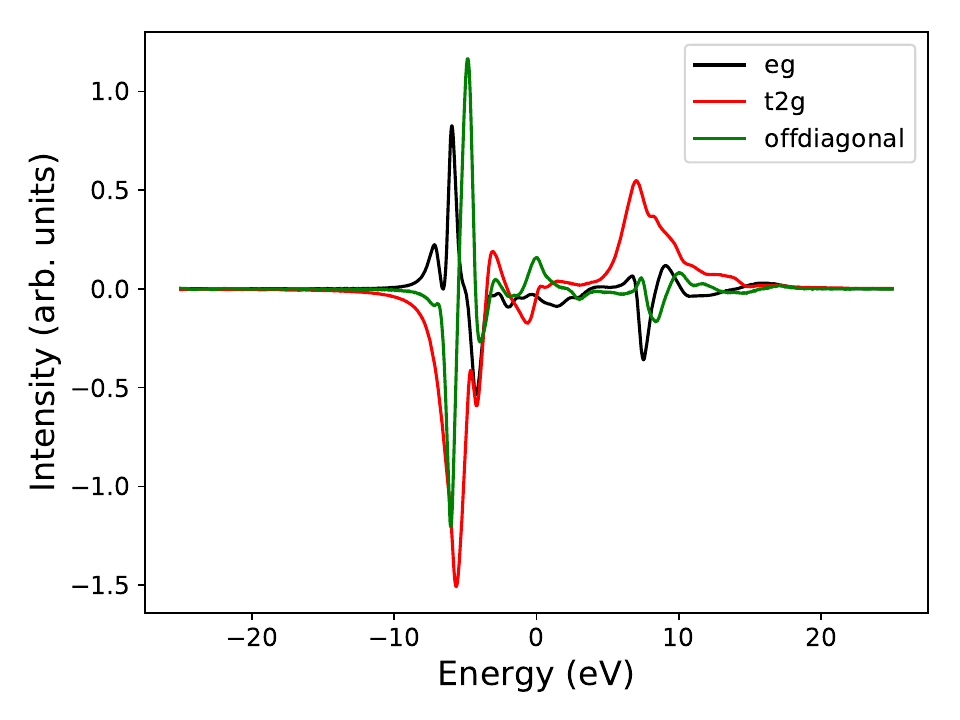}\\
\end{tabular}
}
\caption{Calculated $e_g$ and $t_{2g}$ $L_{2,3}$ XAS (top) and XMCD (bottom) of Fe$_3$O$_4$ (for details see text).
}
\label{fig:Fe3O4-egt2g}
\end{figure}

In this section, we describe the symmetry-decomposed spectra of all compounds investigated here. The symmetry decomposition is made onto $e_g$ and $t_{2g}$ orbitals of the 3d states of the transition metal atom, as described in the main part of this paper. 
The results of the $e_g$ and $t_{2g}$ projections of the XAS and XMCD for Fe$_3$O$_4$ are shown in Fig.~\ref{fig:Fe3O4-egt2g}. In the XAS, we can see that the $t_{2g}$ has a  pronounced shoulder on the right of the main peak at the $L_3$-edge, which in the total spectrum is enhanced due to the off-diagonal elements, while the intensity of the main peak is suppressed. The $e_g$ signal shows a broad peak that begins at the same energy as the $t_{2g}$, but peaks at its shoulder with approximately twice the intensity. At the $L_2$-edge, the contributions from $e_g$ and $t_{2g}$ both show a single peak, where the $t_{2g}$ peak is located at higher energies and keeps the intensity ratio, while the off-diagonal suppresses the signal at the $e_g$ peak and enhances the signal at the $t_{2g}$ peak. 
The $t_{2g}$ peak appears broader because they are the lower-lying orbitals in the lower energy octahedral Fe sites and the higher in the higher energy tetragonal Fe sites. The intensity ratio can be explained by the fact that we can excite into three $t_{2g}$ and two $e_g$ per site.  In the XMCD we see that the first down-pointing peak of the $L_3$-edge is caused by the lower lying $t_{2g}$ orbitals from the octahedral Fe sites that have lower binding energy than the tetragonal site. This signal is further enhanced by the off-diagonal elements. The middle up-pointing peak is not caused by the $e_g$ peaks of the tetragonal site, which are completely compensated by the $t_{2g}$ down-pointing peak, instead, it is caused by the off-diagonal elements.
The $e_g$ and $t_{2g}$ of the two octahedral sites equally contribute to the third peak, which is slightly enhanced by the off-diagonals. The small multiplet peaks between the $L_3$ and $L_2$-edge are caused by a competition between the $t_{2g}$ of the tetragonal and the octahedral sites. The signal at the at the $L_2$-edge is mostly comprised of $t_{2g}$ signal with the $e_g$ and off-diagonals reducing the signal causing the split into two different peaks in the total signal.

\begin{figure}[htp]
\centering{
\begin{tabular}{c}
\includegraphics[scale=0.5]{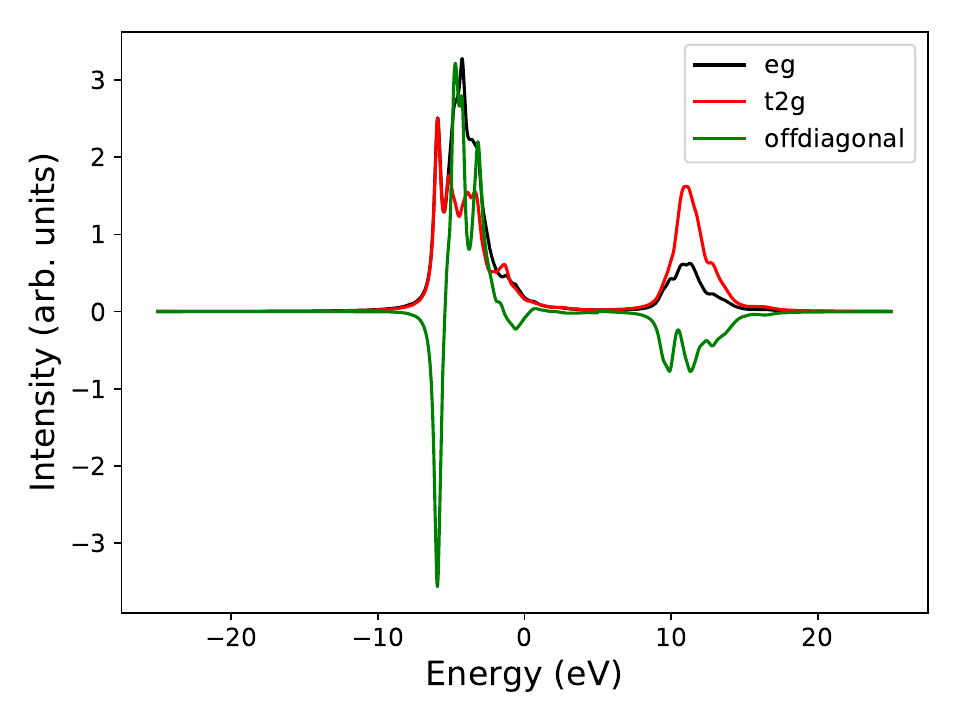}\\
\includegraphics[scale=0.5]{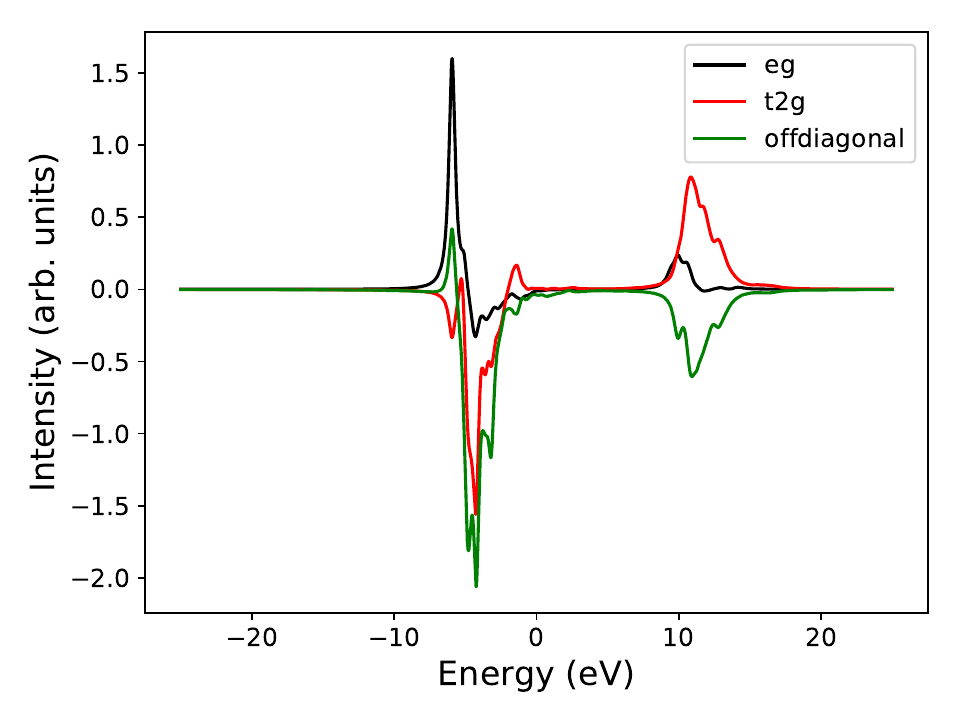}\\
\end{tabular}
}
\caption{Same for Co in CoFe$_2$O$_4$.
}
\label{fig:Co-CoFeOegt2g}
\end{figure}

The  $e_g$ and $t_{2g}$ projected spectra of Co in CoFe$_2$O$_4$ can be seen in Fig.~\ref{fig:Co-CoFeOegt2g}. The first peak in the XAS comes from both the $e_{g}$ and the $t_{2g}$ even though the $t_{2g}$ orbital should be at lower energies, however, they are nearly completely compensated by the off-diagonal elements. After this the $t_{2g}$ drops in value and has three more small peaks, which are overshadowed by the off-diagonals and the $e_g$ signal. The double-peak structure is mainly formed by the off-diagonals and the $e_g$ signals with similar intensities, just that the off-diagonals have a higher intensity at the first peak, while the $e_g$ has the highest intensity at the second peak. The signal at the signal of the $L_2$-edge is dominated by the $t_{2g}$, which is lowered by the off-diagonals. In the XMCD the first peak of the $L_2$-edge is dominated by the $e_g$-signal, while the main peak is mostly caused by the off-diagonal elements followed by the $t_{2g}$. The signal of the $L_2$-edge is dominated by the $t_{2g}$ which is suppressed by the off-diagonals.

\begin{figure}[htp]
\centering{
\begin{tabular}{c}
\includegraphics[scale=0.5]{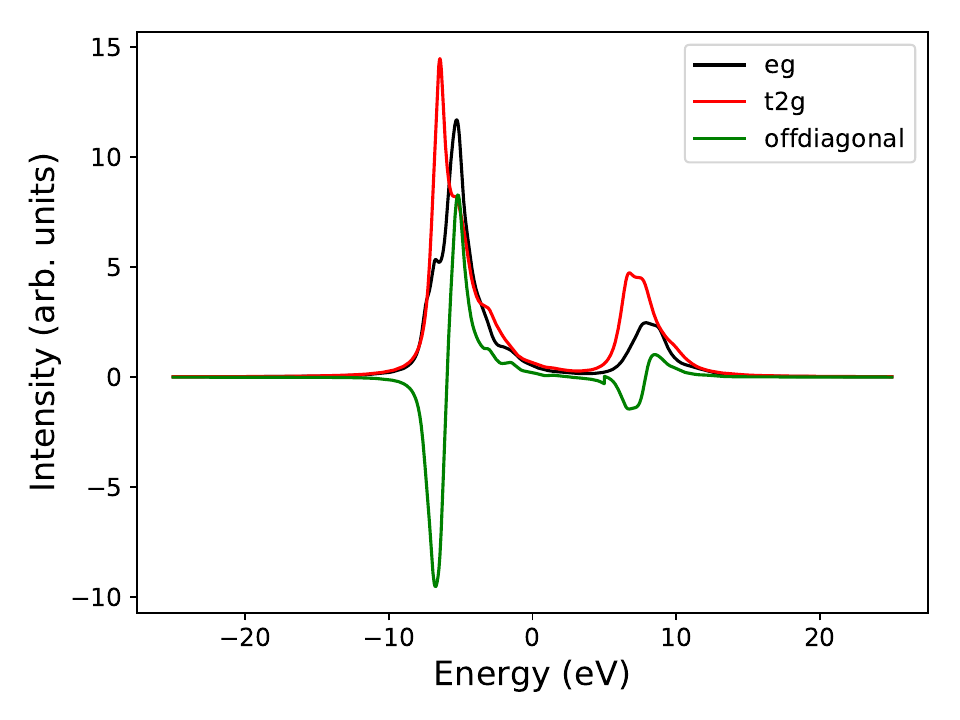}\\
\includegraphics[scale=0.5]{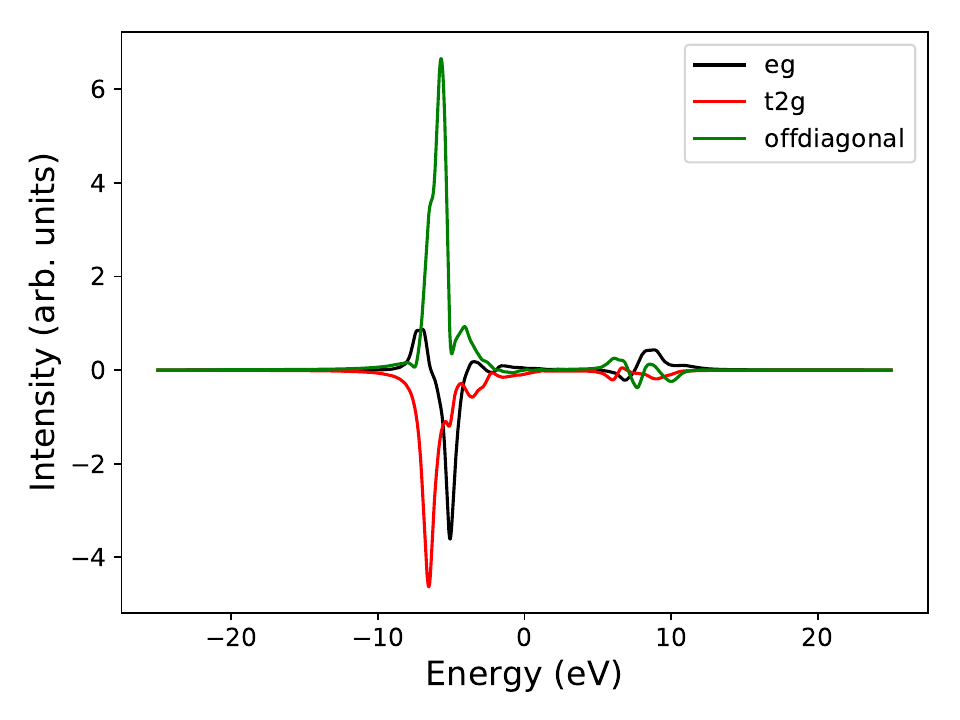}\\
\end{tabular}
}
\caption{Calculated $e_g$ and $t_{2g}$ projection of the Fe $L_{2,3}$ XAS (top) and XMCD (bottom) of CoFe$_2$O$_4$ (for details see text).}
\label{fig:Fe-CoFeOegt2g}
\end{figure}

The Fe-projected $e_g$ and $t_{2g}$ spectra can be seen in Fig.~\ref{fig:Fe-CoFeOegt2g}. The beginning of the $L_3$-edge of the XAS is characterized by the competition of the off-diagonals and the $t_{2g}$ signals, where the $t_{2g}$ signal is slightly bigger than the signal by the off-diagonal terms.
The main peak is mainly caused by the $e_g$ contributions, but the $t_{2g}$ also shows a shoulder here, which has the same intensity as the peak of the off-diagonals.
At the $L_2$-edge, most of the signal is generated by the $t_{2g}$. In the XMCD, we can see that all the peaks of the $L_3$-edge have different origins. The first down peak is caused by the excitations into the $t_{2g}$, the up-pointing peak by the off-diagonals and the last peak by the $e_g$.

\begin{figure}[htp!]
\centering{
\begin{tabular}{c}
\includegraphics[scale=0.5]{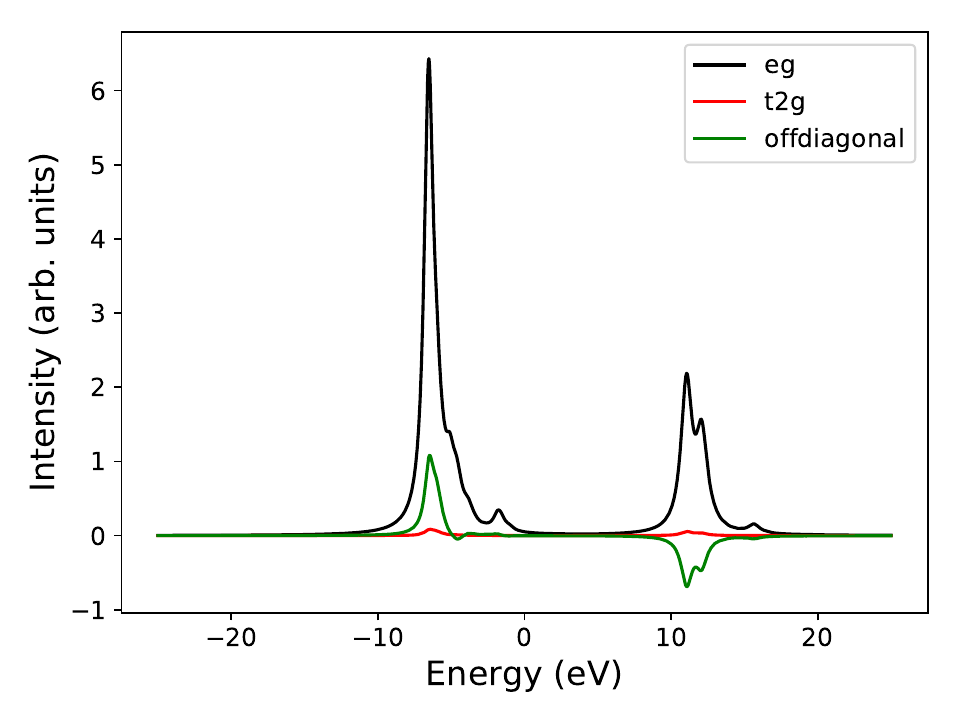}\\
\includegraphics[scale=0.5]{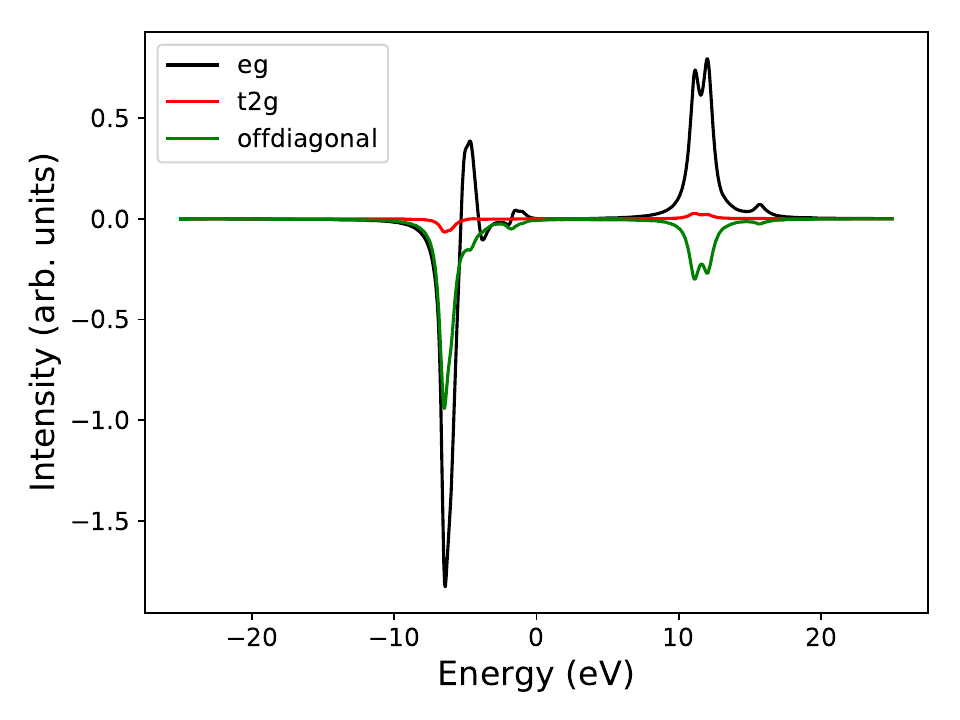}\\
\end{tabular}
}
\caption{Calculated $e_g$ and $t_{2g}$ projection of the Ni $L_{2,3}$ XAS (top) and XMCD (bottom) of NiFe$_2$O$_4$ (for details see text).
}
\label{fig:Ni-NiFeOegt2g}
\end{figure}

 In Fig.~\ref{fig:Ni-NiFeOegt2g} we see the e$_g$- and $t_{2g}$-projected results of the Ni site in NiFe$_2$O$_4$. Here, nearly all the signal comes from the $e_{g}$, because in Ni$^{2+}$ the $t_{2g}$ is fully occupied and one can therefore only make electronic transitions to the $e_g$. 
 
\begin{figure}[htp!]
\centering{
\begin{tabular}{c}
\includegraphics[scale=0.5]{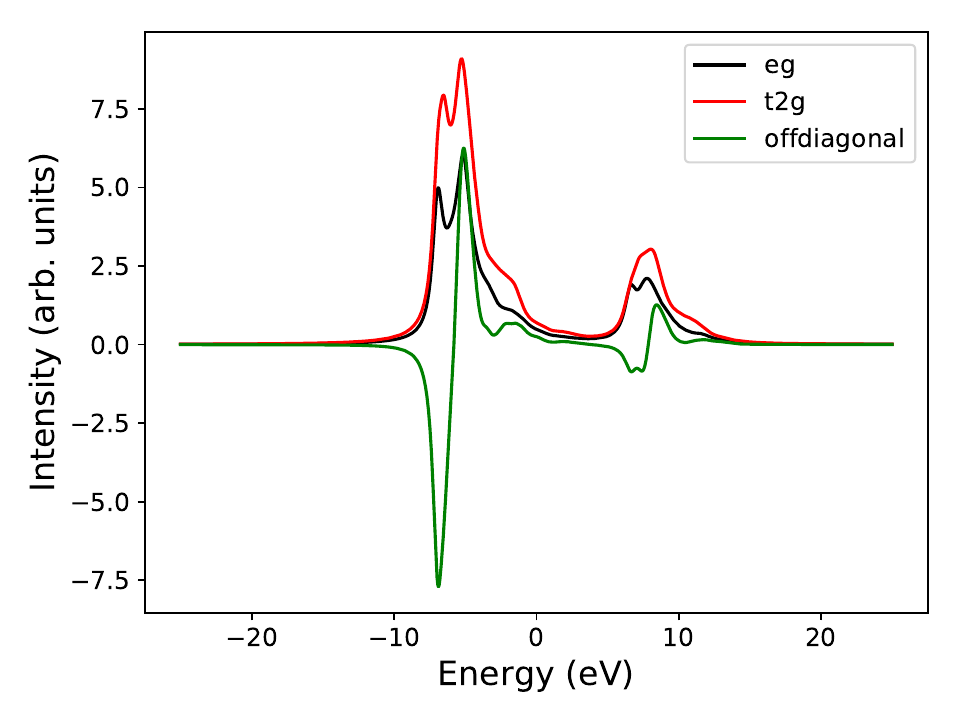}\\
\includegraphics[scale=0.5]{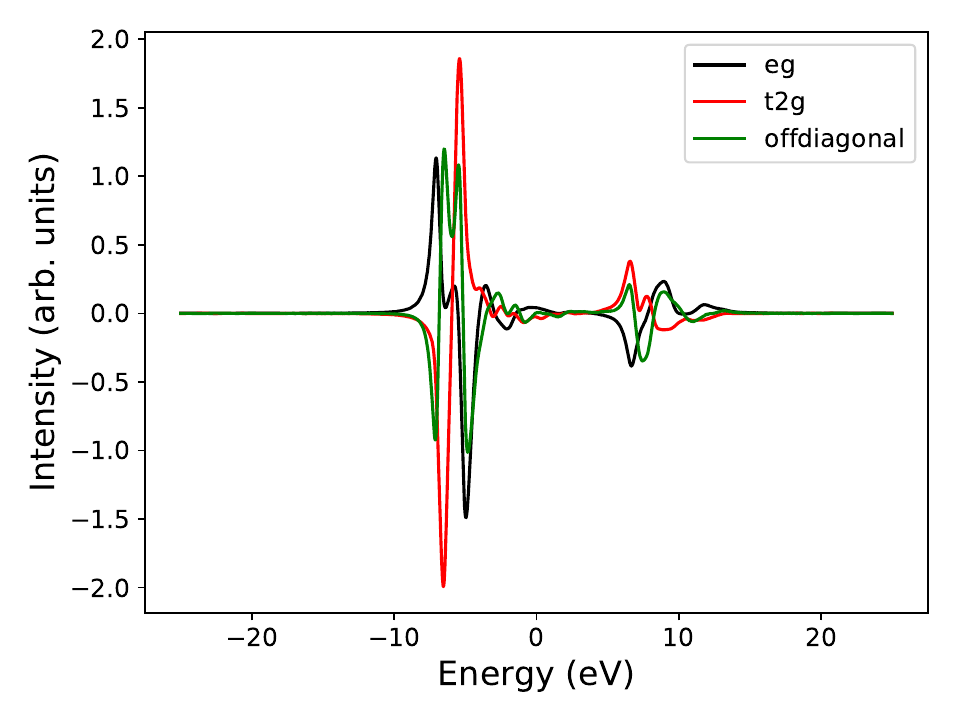}\\
\end{tabular}
}
\caption{Calculated $e_g$ and $t_{2g}$ projection of the Fe $L_{2,3}$ XAS (top) and XMCD (bottom) of NiFe$_2$O$_4$ (for details see text).
}
\label{fig:Fe-NiFeOegt2g}
\end{figure}

 Finally, the result for the Fe sites in NiFe$_2$O$_4$ are shown in 
 Fig.~\ref{fig:Fe-NiFeOegt2g}. In the XAS, we see that the lowest energy peak is caused mostly by the $t_{2g}$, however, the e$_g$ signal also contributes and the off-diagonal elements reduce the intensity. The main peak is caused by a combination of all three signal sources with the biggest contribution coming from the $t_{2g}$. At the $L_2$-edge, one can see that the two peak signals come mostly from the $e_g$ as the $t_{2g}$ is too broad the easily distinguish between the peaks even though it has more intensity. While the off-diagonal reduces the intensity of the first peak and slightly increases the intensity of the second peak.
 The XMCD shows that the first down-pointing peak comes from excitations into the $t_{2g}$ orbitals which corresponds to the lowest lying states of the octahedral sites. At this position, we also have the contributions from the $e_g$ from the tetrahedral site but the intensity is so small that is not visible in the total spectrum. The middle peak that is caused by the tetrahedral site, is instead caused by off-diagonal elements and the $t_{2g}$ contributions.
 The second down-pointing peak is just as expected caused by the higher lying $e_g$ states in the octahedral Fe site. At the $L_2$-edge, we can see that the initial dip in the total spectra is caused by the $e_{g}$ and the off-diagonal signal and the peak after that is caused by the same.
\subsection{DMFT}
\label{appendix:DMFT}
\begin{figure}
    \centering
    \includegraphics[scale=0.5]{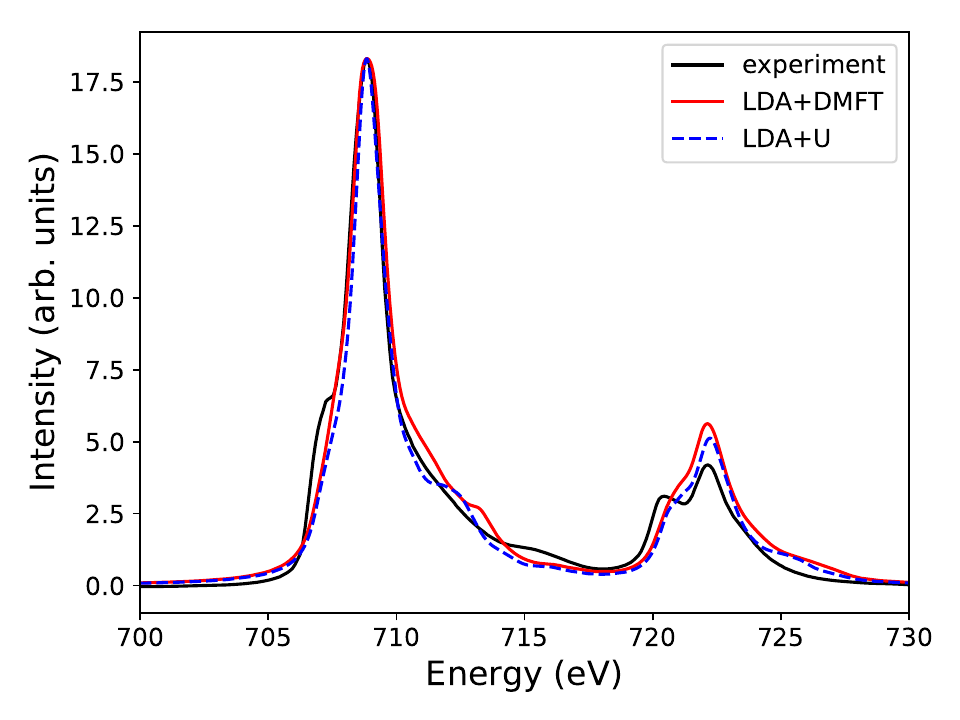}\\
    \includegraphics[scale=0.5]{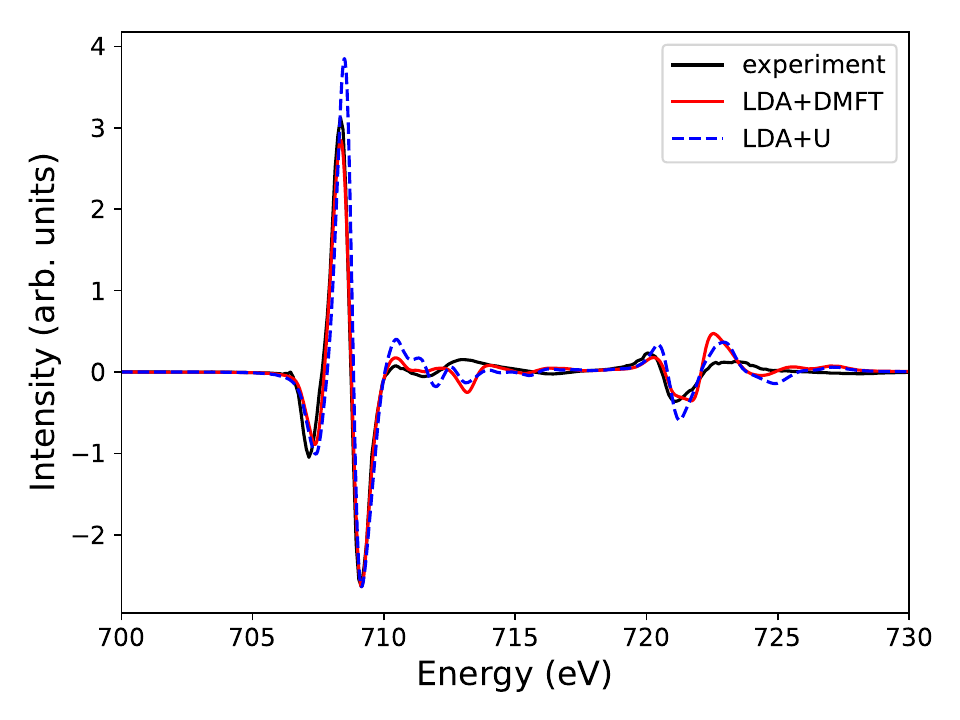}\\
    \caption{ Calculated Fe $L_{2,3}$  XAS (top panel) and XMCD (bottom panel) edges in NiFe$_2$O$_4$ starting from a converged DMFT calculation (red solid) and LDA$+U$ (blue dashed) compared to experiments (black solid).}
    \label{fig:FeDMFT}
\end{figure}

\begin{figure}
    \centering
    \includegraphics[scale=0.5]{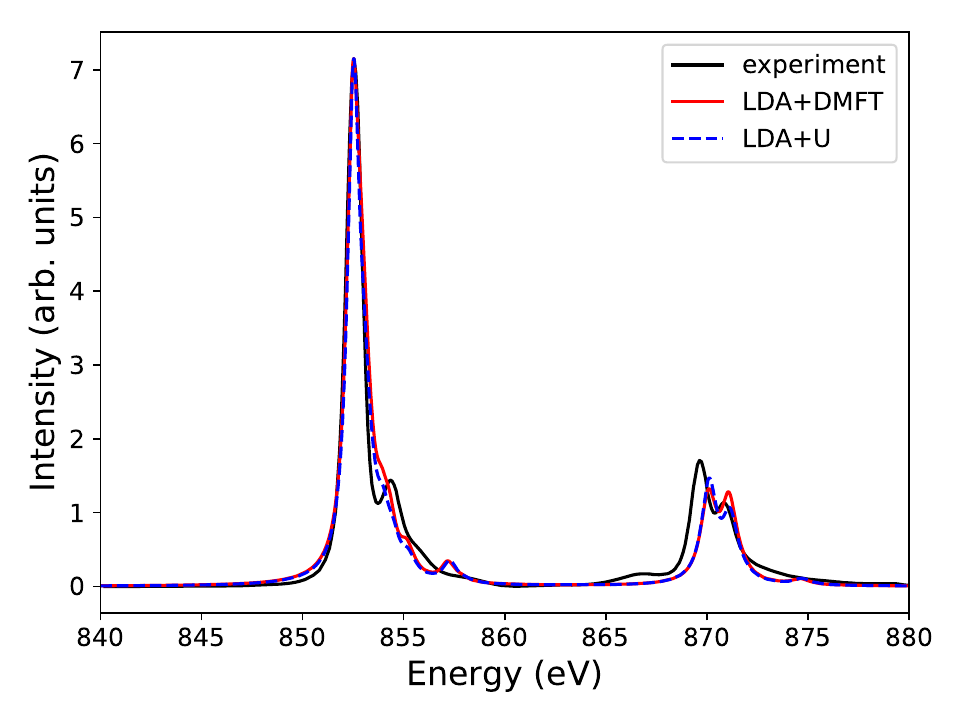}\\
    \includegraphics[scale=0.5]{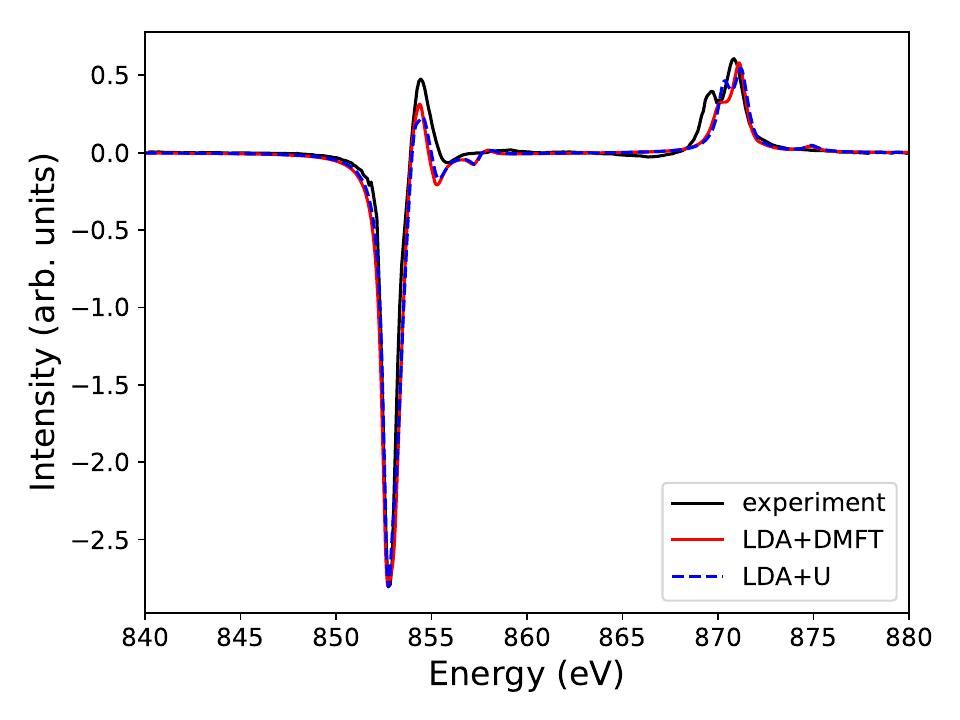}\\
    \caption{Calculated Ni $L_{2,3}$  XAS (top panel) and XMCD (bottom panel) edges in NiFe$_2$O$_4$ starting from a converged DMFT calculation (red solid) and LDA$+U$ (blue dashed) compared to experiments (black solid).}
    \label{fig:NiDMFT}
\end{figure}

In this section, we investigate the effects of using the local Hamiltonian and the hybridization function obtained from a DFT+DMFT calculation, instead of a DFT$+U$ calculation to construct the impurity Hamiltonian of NiFe$_2$O$_4$, while using the same Slater and SOC parameters as before. 
 We used the method described in Ref.~\cite{PhysRevLett.109.186401}, with one bath state per correlated orbital and the ED impurity solver. 
Figs.~\ref{fig:FeDMFT} and \ref{fig:NiDMFT} show the XAS and XMCD spectra of the Fe and Ni sites, respectively, calculated using the same parameters as  Figs.~\ref{fig:NiFeO-Fe} and \ref{fig:NiFeO-Ni} except that the relative corelevel shift $\Delta \epsilon_p$ was adjusted to the values extracted from the DFT+DMFT calculation.
Here, we can see that the spectra are very similar. In the case of the Fe-projected XAS, we can see that the spectrum looks broader, which causes the $L_2$-edge to be even more overestimated. In the XMCD, we can see that the second peak and the oscillations after the third peak of the $L_3$-edge are better reproduced in the DFT+DMFT calculation. Similarly, we can see improvements at the beginning of the $L_2$-edge, while the end is even more overestimated than in the DFT+$U$ calculation.
In the Ni-projected XAS, we can see that the $L_3$-edge shoulder is more strongly pronounced in the DFT+DMFT calculation. At the $L_2$-edge however, we can see a double peak structure in the DFT+DMFT solution instead of a pronounced first peak which we see in the experiment and DFT+$U$ calculation. In the XMCD, we can see that the DFT+DMFT calculation reproduces the experiment even better than the DFT+$U$ calculation with a more pronounced $L_3$-edge shoulder and a better relative intensity of the peaks in the $L_2$-edge.
The similarities between the results from DFT$+U$ and DFT+DMFT are expected, because the sites mostly hybridize to the O sites, which are only treated using DFT. Another reason for the similarities is that we included all exchange interaction in the $+U$-term of the Hamiltonian (the so-called LDA$+U$ approximation, which is not identical to the LSDA$+U$ approximation, where exchange splitting is also included in the density functional), that guarantees that exchange interactions are treated in the same way in the two approaches. 
 

\clearpage
\bibliography{References.bib}
\bibliographystyle{apsrev4-1}
\end{document}